\documentclass{article}

\usepackage{arxiv}

\usepackage[utf8]{inputenc} % allow utf-8 input
\usepackage[T1]{fontenc}    % use 8-bit T1 fonts
\usepackage{hyperref}       % hyperlinks
\usepackage{url}            % simple URL typesetting

\usepackage{amsbsy}
\usepackage{amsfonts}       % blackboard math symbols
\usepackage{amsfonts}
\usepackage{amsmath}
\usepackage{amssymb}
\usepackage{amsthm}
\usepackage{booktabs}       % Professional-quality tables.
\usepackage{caption}        % Table title text edition.
\usepackage{nicefrac}       % compact symbols for 1/2, etc.
\usepackage{microtype}      % microtypography
\usepackage{multirow}       % Enables multirow command.
\usepackage{relsize}        % Larger/smaller math symbols.

\usepackage{lipsum}	  % Can be removed after putting your text content
\usepackage{graphicx}
\usepackage{natbib}
\usepackage{doi}

\newtheorem{definition}{Definition}
\newtheorem{theorem}{Theorem}

\title{Semiparametric modeling for multivariate survival data via copulas}

\author{ 
{
\hspace{1mm}Walmir D. R.~Miranda Filho}
\\
Department of Statistics\\
Federal University of Minas Gerais\\
Belo Horizonte, MG, Brazil \\
\texttt{walmir.reis.miranda.filho@gmail.com} \\
\And
{
\hspace{1mm}Fábio N.~Demarqui} \\
Department of Statistics\\
Federal University of Minas Gerais\\
Belo Horizonte, MG, Brazil \\
\texttt{fndemarqui@gmail.com} \\
}

\renewcommand{\headeright}{}
\renewcommand{\undertitle}{}
\renewcommand{\shorttitle}{Semiparametric modeling for multivariate survival data via copulas}

\hypersetup{
pdftitle={Semiparametric modeling for multivariate survival data via copulas},
pdfsubject={q-bio.NC, q-bio.QM},
pdfauthor={Walmir D. R.~Miranda Filho, Fábio N.~Demarqui},
pdfkeywords={Archimedean copulas, Marginal survival functions, Baseline distributions, Regression model classes},
}

\begin{document}
\maketitle

\begin{abstract}
	We propose a new class of multivariate survival models based on archimedean copulas with margins modeled by the Yang and Prentice (YP) model. The Ali-Mikhail-Haq (AMH), Clayton, Frank, Gumbel-Hougaard (GH), and Joe copulas are employed to accommodate the dependency among marginal distributions. Baseline distributions are modeled semiparametrically by the piecewise exponential (PE) distribution and the Bernstein polynomials. The new class of models possesses some attractive features: i) the ability to take into account survival data with crossing survival curves; ii) the inclusion of the well-known proportional hazards (PH) and proportional odds (PO) models as particular cases; iii) greater flexibility provided by the semiparametric modeling of the marginal baseline distributions; iv) the availability of closed-form expressions for the likelihood functions, leading to more straightforward inferential procedures. We conducted an extensive Monte Carlo simulation study to evaluate the performance of the proposed model. Finally, we demonstrate the versatility of our new class of models through the analysis of survival data involving patients diagnosed with ovarian cancer.
\end{abstract}

% keywords can be removed

\section{Introduction}

Survival analysis is a wide field is Statistics that deals with time data ranging from a well-defined start point until the occurrence of a particular event of interest, taken as the endpoint criteria, for a set of subjects under study \citep{Coll2015}. For instance, in medical research the start point can correspond to the recruitment of a subject diagnosed with a disease, and the event of interest can be the death, or the appearance or recurrence of a symptom. An important feature present in many survival analysis studies is right-censoring \citep{Coll2015}. An observed time is said to be right-censored if the event of interest has not been observed for a subject, but it is known that the subject has survived up to the observed time. A right-censored time can arise if a subject has been lost to follow-up before the event occurrence (for example, a patient that moves to another country and can no longer be tracked), the event has not been observed at the end of the study for a subject, or death has been caused for a reason unrelated to the event of interest. A crucial assumption that will be made from now on is that the actual survival time, whether it is observed or not, does not depend on any censoring mechanism. Then, censoring is said to be independent, operating randomly over all subjects under study.

Depending on the study, survival data can be clustered by a grouping variable or even by subjects. This last can occur when the event of interest is observed more than one time for the same subject, and for all subjects under study. Thus, each cluster is composed by two or more observed times. Since data from the same cluster are expected to be correlated, it cannot be modeled as if its observed times were independent. A theoretical and well-suited framework to model clustered data is the copula, a multidimensional distribution function over a fixed number of standard uniform univariate margins, equal to the size of each cluster. Copulas can link marginal survival functions to form a joint survival distribution to model clustered data \citep{MaRa2019}. Not only survival or distribution functions but any function restricted to the unit interval can be used as a marginal component of a copula model.

The first investigation of a multivariate standardized distribution function was made by \cite{Hoef1940}, who worked with bivariate standardized distributions with uniform margins on the interval $[-1/2, 1/2]$. As \cite{Schw1991}, \emph{apud} \cite{Nels2006}, pointed out, Hoeffding could have discovered copulas if he had chosen the unit square $[0,1]^2$ instead. \cite{Fero1956} considered standardized distribution functions defined on the unit cube $[0,1]^3$, but the term ``copula'' (a Latin word for link) was introduced by \cite{Skla1959}. He also established a result connecting multivariate distribution functions and their one-dimensional standard uniform margins, posteriorly known as the Sklar's Theorem \citep{Nels2006}. Until the early 1980s, results for the copula theory were mostly limited in the context of probabilistic metric spaces. At that time, \cite{ScWo1981} studied different criteria for measures of dependence and concluded that copulas provide a tool to analyze the dependence among random variables. They also showed that copulas are invariant under strictly increasing transformations. Thereby, it was proved that copulas can resume information on the dependence structure among random variables. Some recent references on copula theory and applications are the books of \cite{Nels2006}; \cite{HJoe2015}; \cite{DuSe2015}; Flores \emph{et al.} \citeyearpar{Flor2017} and Hofert \emph{et al.} \citeyearpar{Hofe2018}.

Once copula margins are defined on the unit interval $[0,1]$, they could be represented each one by a survival function, forming a joint survival distribution. The first use of copulas as multivariate survival models dates back to \cite{Clay1978}. He noted that, when adjusting for covariates, marginal survival functions and the copula dependence parameter can unveil underlying factors influencing the probability of event times simultaneously. \cite{Oake1982}, in turn, reparameterized Clayton's model and proposed corrections for his likelihood expression and variance estimation. The Clayton copula is a member of a wide class used in survival analysis: the Archimedean copulas, which also includes the Ali-Mikhail-Haq (AMH), Frank, Gumbel-Hougaard (GH), and Joe copulas (see \cite{Nels2006} and references therein).

Marginally, the survival function can be modeled through a baseline distribution combined with a regression model class, if there is any covariate. The baseline distribution can be a parametric model family, such as the Weibull distribution \citep{Weib1951}, or a parametric model with a nonparametric appeal (without fixing the number of parameters and therefore its functional form), such as the Bernstein Polynomials (BP) from \cite{Bern1913} or the Piecewise Exponential (PE) proposed by \cite{KaPr1973}. Combining one of the later with a regression model class yields the called semiparametric models. The traditional Cox's Proportional Hazards (PH) model \citep{DCox1972} is an example of semiparametric model: it takes advantage of the proportionality assumption for the hazards ratios over time to estimate directly only the regression parameters, dropping its baseline (hazard) function. However, not always such assumption is verified for the survival data, and in that case some alternatives of regression model classes were developed in the literature, such as the Proportional Odds (PO) model proposed by \cite{Benn1983} and the Yang-Prentice (YP) model from \cite{YaPr2005}. The PO class has a similar approach to the PH class, but adopts the assumption of proportionality on the odds ratios over time. In its turn, the YP class arose as a more flexible model to accommodate crossing survival curves, which can appear (for instance) when a treatment is effective in the long run but may present adverse effects to the subjects in early stages. Containing the PH and PO models as particular cases, the YP class can be used for model-fitting checking and may provide a more accurate inference when either the assumption of proportional hazards or odds are violated.

As examples of survival models as marginal components of an Archimedean copula, under independent right-censoring, there are the works of Goethals \emph{et al.} \citeyearpar{Goet2012}, who fitted copula models with Weibull PH components for bivariate clustered data and compared it with shared frailty models; Louzada \emph{et al.} \citeyearpar{Louz2013} for bivariate copulas with survival mixture models on each of their components; Prenen \emph{et al.} \citeyearpar{Pren2017} for multivariate clusters with variable size, and \cite{MaRa2019} for additive copula models with monotonic splines. However, regression model classes of practical importance such as the aforementioned PO and YP classes, or flexible baseline distributions like the BP and PE models, still have not been addressed to joint survival function modeling through Archimedean copulas.

To evaluate the fitting of bivariate Archimedean copula models with marginal survival functions (henceforward called survival copula models), this work focus on specifying one among a variety of baseline distributions (Weibull, BP or PE), combined with a regression model class (PH, PO or YP), for each Archimedean copula (AMH, Clayton, Frank, GH or Joe). The computational framework consists of programmed functions in Stan (\citeyear{Stan2020}), an open-source language designed to define custom likelihood functions. %and, for a Bayesian analysis, prior specifications.
The present paper uses the integrated version of Stan with R, the \texttt{rstan} (\citeyear{RStan2020}), to maximize the log-likelihood function for the proposed survival copula models. %Under a frequentist approach of inference, likelihood maximization is done through the Broyden-Fletcher-Goldfarb-Shanno (BFGS) algorithm \citep{NoWr2006}. %Under a full Bayesian inference approach, Stan draws posterior samples through the Hamiltonian Monte Carlo (HMC) algorithm extended with the No-U-Turn Sampler, preventing a random walk path and greatly reducing the sensitivity to correlated parameters, when compared to traditional Markov Chain Monte Carlo methods, such as random walk Metropolis and Gibbs Sampling. Thereby, their chains converge faster to the target distribution \citep{HoGe2014}.

The main contribution of the present paper is to provide a novel and extended theoretical framework on a fully likelihood-based approach to handle clustered survival data, under independent right-censoring and allowing a wide range of behaviors for marginally fitted survival functions. This is done mainly through the choice of flexible models with a nonparametric appeal for the baseline distribution (\emph{i.e.}, their number of parameters is not always the same, like Weibull models) combined with the YP regression model class, which allows crossing survival times while also containing the PH and PO classes as special cases (if the proportional hazards or odds ratios assumption are reasonable). %It should be emphasize that a survival (Archimedean) copula modeling that marginally specifies BP or PE models as baseline distributions and/or PO or YP regression model classes has not been seen in the literature.
% An R package called copSurv is being developed to cover fitting and analysis for all proposed models.

Some substantial advantages of the proposed methodology here are: (i) for a frequentist approach, copulas do not require integrating out the dependence parameter in order to estimate it; (ii) for some Archimedean copulas, it is possible to model negative dependence; %(iii) for a Bayesian approach, copulas do not impose restrictions on the dependence parameter distribution moments to ensure identifiability;
(iii) when marginally specifying the YP regression model class, copulas provide a joint survival distribution linking marginal continuous survival functions that allows the detection of their corresponding intersection points (\emph{i.e.}, the marginal crossing survival times); (iv) the flexibility of marginal baseline distributions with a nonparametric appeal allows a great variety of shapes for the hazard function; (v) the routine to implement the
(fully) maximum likelihood estimation and inference is straightforward and does not demand intermediate steps.

This work is organized as follows. The proposed survival copula modeling is described in Section 2. An extensive Monte Carlo simulation to evaluate the performance of survival copula models is done in Section 3. A data collection of patients with ovarian cancer from Ganzfried \emph{et al.} \citeyearpar{Ganz2013} is used as a real application to fit the proposed models in Section 4. Finally, main conclusions, final remarks and a discussion on future research are presented in Section 5.

\section{Survival Copula Modeling Formulation}

As stated in Section 1, copulas are a way to model dependence among observed values from the same cluster (subject). In the following subsections, the main topics of copula theory are covered, focusing on their characterization as joint survival functions. %A measure to evaluate the strength of dependence given the estimate for a copula parameter is also discussed. 
For the development below, each copula margin is supposed to be an absolutely continuous function associated to a random variable with continuous support. A more general version, including random variables with discrete support, is treated by \cite{Nels2006}.

\subsection{Copula Theory}

In the literature, copulas are generally referred to as ``functions that assemble or couple one-dimensional distribution functions to a multivariate distribution function'' or as ``distribution functions whose one-dimensional margins are all uniform'' \citep{Nels2006}. Not all multivariate distributions are a copula itself, but each one can be reparameterized to a copula (see Section 1.2 of \cite{HJoe2015} for some examples).
\begin{definition}[Flores \emph{et al.}, \citeyear{Flor2017}, p. 2-3]\label{def3.1}
	Let $d$ be a positive integer value and $I = [0,1]$ be the unit interval. A $d$-dimensional copula is a function $C:I^d \mapsto I$ that attends the following properties:
	
	\noindent (a) $C(u_1, \ldots, u_{j - 1}, 0, u_{j + 1}, \ldots, u_d) = 0$ for all $j = 1, \ldots, d$;
	
	\noindent (b) $C(1, \ldots, 1, u_j, 1, \ldots, 1) = u_j$ for all $j = 1, \ldots, d$;
	
	\noindent (c) For each rectangle $[\boldsymbol{a}, \boldsymbol{b}] = \prod_{j = 1}^d [a_j, b_j] \subset I^d$, with $a_j \leqslant b_j$ for all $j = 1, \ldots, d$ in the Cartesian product of intervals on the right-hand side, its volume is given by
	$$
	\text{Vol}_C([\boldsymbol{a}, \boldsymbol{b}]) = \sum_{j_1 = 1}^2 \cdots \sum_{j_d = 1}^2 (-1)^{j_1 + \cdots + j_d} C\left(u_{1_{j_1}}, \ldots, u_{d_{j_d}}\right) \geqslant 0,
	$$
	where $u_{j_1} = a_j$ and $u_{j_2} = b_j$ for all $j = 1, \ldots, d$.
\end{definition}
A fundamental result established by \cite{Skla1959} allowed a representation for the joint distribution of $d$ random variables as a function of marginal distribution functions. Let $F_1, \ldots, F_d$ be a set of continuous distribution functions with range (\emph{i.e.}, all values that $F_j$ can assume) $\text{ran}F_j = I, j = 1, \ldots, d$, and $F_j^{-1}$ the corresponding inverse function. Then
\begin{theorem}[Sklar's Theorem, \citeyear{Skla1959}, \emph{apud} Hofert \emph{et al.}, \citeyearpar{Hofe2018}, p. 23]\label{t3.2St}
	Let $F$ be the continuous joint distribution function of a $d$-dimensional random vector $\boldsymbol{X} = (X_1, \ldots, X_d)$ with marginal continuous distribution functions $F_1, \ldots, F_d$. Then, there exists a $d$-dimensional copula $C$ such that	
	\begin{equation}\label{eq3.2jdfascop}
	F(\boldsymbol{x}) = C\left[F_1(x_1), \ldots, F_d(x_d)\right], \quad \boldsymbol{x} = (x_1, \ldots, x_d) \in \mathbb{R}^d.
	\end{equation}
	Moreover, $C$ is uniquely defined by the Cartesian product of the $d$ ranges, $\prod_{j=1}^{d} \text{ran}F_j = I^d$, and there given by
	\begin{equation}\label{eq3.3copgen}
	C(\boldsymbol{u}) = F\left[F_1^{-1}(u_1), \ldots, F_d^{-1}(u_d)\right], \quad \boldsymbol{u} \in I^d.
	\end{equation}
	Conversely, given a $d$-dimensional copula $C$ and a sequence of univariate distribution functions $F_1, \ldots, F_d$, $F$ defined in \eqref{eq3.2jdfascop} is a $d$-dimensional joint distribution function.
\end{theorem}

%The second result in \eqref{eq3.3copgen} provides a method for construction of copulas from joint distributions, and from \eqref{eq3.2jdfascop} it follows that $F$ is absolutely continuous if and only if all marginal distribution functions $F_1, \ldots, F_d$ and the copula $C$ are absolutely continuous. To attend this property, the copula should admit a density $c$ such that
%\begin{equation}\label{eq3.3cdf}
%c(\boldsymbol{u}) = \dfrac{\partial^d}{\partial u_d \cdots u_1} C(u_1, \ldots, u_d), \boldsymbol{u} \in I^d
%\end{equation}
%exists and is integrable. Being the copula itself a multivariate distribution function, each successive partial derivative with respect to one or more margins is a joint conditional distribution function for the remaining copula margins. Differentiating $d$ times, expression \eqref{eq3.3cdf} is obtained. For the bivariate case ($d = 2$), the first-order derivative $\partial C(u_1, u_2) / \partial u_1$ is the conditional distribution for the margin $U_2$ given that $U_1 = u_1$, also denoted by $C_{2|1}(u_2|u_1)$. Similarly, $\partial C(u_1, u_2) / \partial u_2 = C_{1|2}(u_1|u_2)$ is the conditional distribution for the margin $U_1$ given that $U_2 = u_2$. Finally, $\partial^2 C(u_1, u_2) / \partial u_2 \partial u_1 = c(u_1, u_2)$ is the density function associated with the copula $C$.

From Theorem \ref{t3.2St}, a similar relationship can be constructed for a joint survival function starting from marginal survival functions instead of the distribution ones, since the survival function is defined as the complement of a distribution function. In fact, the Sklar's Theorem can be formulated using survival functions as well.
\begin{theorem}[Sklar's Theorem for Survival Functions, Hofert \emph{et al.}, \citeyearpar{Hofe2018}, p. 41]\label{t3.3St}
	Let $S$ be the joint survival function of a $d$-dimensional random vector $\boldsymbol{T} = (T_1, \ldots, T_d)$ with marginal continuous survival functions $S_1, \ldots, S_d$. Then, there exists a $d$-dimensional survival copula $\overline{C}$ such that
	\begin{equation}\label{eq3.6jsfascop}
	S(\boldsymbol{t}) = \overline{C}\left[S_1(t_1), \ldots, S_d(t_d)\right], \quad \boldsymbol{t} = (t_1, \ldots, t_d) \in \mathbb{R}^d.
	\end{equation}
	The survival copula $\overline{C}$ is uniquely defined by the Cartesian product of the $d$ ranges, $\prod_{j=1}^{d} \text{ran}S_j = I^d$, and given by
	\begin{equation}\label{eq3.7copgen}
	\overline{C}(\boldsymbol{u}) = S\left[F_1^{-1}(1 - u_1), \ldots, F_d^{-1}(1 - u_d)\right], \quad \boldsymbol{u} \in I^d.
	\end{equation}
	Conversely, given a $d$-dimensional survival copula $\overline{C}$ and a sequence of univariate survival functions $S_1, \ldots, S_d$, $S$ defined in \eqref{eq3.6jsfascop} is a $d$-dimensional survival function.
\end{theorem}
The survival copula $\overline{C}$ in \eqref{eq3.7copgen} is also a distribution function: note that for all $j \in 1, \ldots, d, u_j = S_j(t_j) = 1 - F_j(t_j)$. Since $F_j$ and $S_j$ are absolutely continuous for all $j$, then $F_j^{-1}(1 - u_j) = S_j^{-1}(u_j) = t_j$, and
\begin{align*}
\overline{C}(\boldsymbol{u}) &= S\left[F_1^{-1}(1 - u_1), \ldots, F_d^{-1}(1 - u_d)\right] = S\left[S_1^{-1}(u_1), \ldots, S_d^{-1}(u_d)\right] \\
&= \mathbb{P}\left[S_1^{-1}(U_1) \geqslant S_1^{-1}(u_1), \ldots, S_d^{-1}(U_d) \geqslant S_d^{-1}(u_d)\right] \\
&= \mathbb{P}\left[S_1(T_1) \leqslant S_1(t_1), \ldots, S_d(T_d) \leqslant S_d(t_d)\right] \\
&= \mathbb{P}\left[U_1 \leqslant u_1, \ldots, U_d \leqslant u_d\right].
\end{align*}
Now, let $i \in \{1, \ldots, n\}$ be the cluster (subject) index and $\boldsymbol{y}_i$ be a $d$-dimensional vector of observed times. Suppose that all marginal times $y_{i;1}, \ldots, y_{i;j}$, $j = 1, \ldots, d$ follow the same univariate parametric survival model with a set of parameters $\boldsymbol{\kappa}_1, \ldots, \boldsymbol{\kappa}_d$ for each margin, fixed for all $i$. Let $d_i = \sum_{j=1}^{d} \delta_{i;j}$ be the quantity of survival times, where $\delta_{i;j}$ is the censoring indicator random variable value for $y_{i;j}$. Following p. 96--97 of \cite{DuJa2007} and p. 486 of Prenen \emph{et al.} \citeyearpar{Pren2017}, but supposing that all clusters have the same size $d$, the contribution of each cluster $i$ to the likelihood function for the survival copula model in \eqref{eq3.6jsfascop} is given by
\begin{align}
L_i(\theta, \boldsymbol{\kappa}_1, \ldots, \boldsymbol{\kappa}_d | \boldsymbol{y}_i)
&\propto (-1)^{d_i} \dfrac{\partial^{d_i}}{\partial \prod_{j=1}^d (y_{i;j})^{\delta_{i;j}}} S(\boldsymbol{y}_i | \theta, \boldsymbol{\kappa}_1, \ldots, \boldsymbol{\kappa}_d) \nonumber \\
&= \dfrac{\partial^{d_i}}{\partial \prod_{j=1}^d (y_{i;j})^{\delta_{i;j}}} \overline{C}\left[S_1(y_{i;1}|\boldsymbol{\kappa}_1), \ldots, S_d(y_{i;d} | \boldsymbol{\kappa}_d) | \theta\right] \prod_{j=1}^d f_j(y_{i;j} | \boldsymbol{\kappa}_j)^{\delta_{i;j}}, \label{eq3.6copllcontrib}
\end{align}
Thus, the survival copula likelihood function is given by the product over all subjects $i$, $i = 1, \ldots, n$, of expression \eqref{eq3.6copllcontrib}. If $d=2$, it is expressed as
\begin{align}
L_{\overline{C}}(\theta, \boldsymbol{\kappa}_1, \boldsymbol{\kappa}_2 | \boldsymbol{y})
&\propto \prod_{i=1}^{n} \left\{\overline{C}\left[S_1(y_{i;1} | \boldsymbol{\kappa}_1), S_2(y_{i;2} | \boldsymbol{\kappa}_2) | \theta\right]\right\}^{(1 - \delta_{i;1})(1 - \delta_{i;2})} \nonumber \\
&\times \left\{\dfrac{\partial \overline{C}\left[S_1(y_{i;1} | \boldsymbol{\kappa}_1), S_2(y_{i;2} | \boldsymbol{\kappa}_2) | \theta\right]}{\partial y_{i;1}}\right\}^{\delta_{i;1}(1 - \delta_{i;2})} % \nonumber \\
% &\times
\left\{\dfrac{\partial \overline{C}\left[S_1(y_{i;1} | \boldsymbol{\kappa}_1), S_2(y_{i;2} | \boldsymbol{\kappa}_2) | \theta\right]}{\partial y_{i;2}}\right\}^{(1 - \delta_{i;1}) \delta_{i;2}} \nonumber \\
&\times \left\{\dfrac{\partial^2 \overline{C}\left[S_1(y_{i;1} | \boldsymbol{\kappa}_1), S_2(y_{i;2} | \boldsymbol{\kappa}_2) | \theta\right]}{\partial y_{i;1} \partial y_{i;2}}\right\}^{\delta_{i;1} \delta_{i;2}} f_1(y_{i;1} | \boldsymbol{\kappa}_1)^{\delta_{i;1}} f_2(y_{i;2} | \boldsymbol{\kappa}_2)^{\delta_{i;2}}.%,
\end{align}
%which is a product involving the survival copula $\overline{C}(u_1, u_2)$ in the first line; the conditional distribution functions $\overline{C}_{2|1}(u_2 | u_1)$ and $\overline{C}_{1|2}(u_1 | u_2)$ in the second line, respectively, and the joint density function $\overline{c}(u_1, u_2)$, multiplied by the marginal densities, in the third line.

Initially, inferential procedures in survival copula models were made in two stages, first estimating the marginal survival and density functions, and then maximizing the likelihood function after replacing the copula function and its derivatives by their estimated versions from the first stage \citep{ShLo1995}. This is similar to adopt a profile likelihood estimation, treating $\theta$ as a nuisance parameter. In this work, however, the survival copula likelihood will be maximized at once, estimating all parameters from copula and marginal survival functions simultaneously. Thus, all covariance structure among parameters is incorporated in the inference, avoiding underestimation on their standard errors.

\subsubsection{Archimedean Copulas}

According to \cite{Nels2006}, the class of Archimedean copulas has a broad range of applications due to its easy construction and attractive mathematical properties. Originally, the class appeared as part of the development of a probabilistic version for the triangle inequality \citep{Schw1991}. Archimedean copulas are constructed through an additive generator function and its pseudo-inverse.
\begin{definition}[Additive Generator, \citealt*{DuSe2015}, p. 196]
	A function $\psi : [0, \infty) \mapsto I$, $I = [0, 1]$, is said to be additive generator if:
	
	\noindent (a) $\psi$ is continuous and decreasing;
	
	\noindent (b) $\psi(0) = 1$ and $\lim\limits_{w \rightarrow \infty} \psi(w) = 0$;
	
	\noindent (c) $\psi$ is strictly decreasing on the interval $[0, w_0]$, where $w_0 = \inf\{w > 0 : \psi(w) = 0\}$.
\end{definition}
Unless when said otherwise, a generator $\psi$ is always meant to be an additive generator. Its pseudo-inverse, denoted as $\psi^{(-1)}(w)$, is equal to the inverse function $\psi^{-1}(w)$ if $w \in (0,1]$ and equal to $w_0$ if $w = 0$.
\begin{definition}[Archimedean Copula, \citealt*{DuSe2015}, p. 196]
	A $d$-dimensional copula is said to be Archimedean if
	\begin{equation}\label{eq3.9arccop}
	C(\boldsymbol{u}) = \psi\left[\psi^{(-1)}(u_1) + \ldots + \psi^{(-1)}(u_d)\right] = \psi\left[\sum_{j=1}^d \psi^{(-1)}(u_j)\right], \quad \boldsymbol{u} \in I^d.
	\end{equation}
\end{definition}
Since Archimedean copulas are defined under $\psi$, the notation $C_\psi$ can replace $C(\boldsymbol{u})$ in \eqref{eq3.9arccop}. Not all generators are suitable to construct an Archimedean copula: a generator $\psi$ needs also to be a $d$-monotone function for some value $d$.
\begin{definition}[%$d$-Monotone and Completely Monotone Functions
$d$-monotonicity, \citealt*{DuSe2015}, p. 197]
	A function $\psi : (a, b) \mapsto \mathbb{R}$ is said to be $d$-monotone in $(a, b)$, where $-\infty \leqslant a < b \leqslant +\infty$ and $d \geqslant 2$, if:
	
	\noindent (a) $\psi$ admits derivatives $\psi^{(k)}$ up to the order $k = d - 2$;
	
	\noindent (b) For all $w \in (a, b)$, $(-1)^k \psi^{(k)}(w) \geqslant 0$, for $k \in \{0, 1, \ldots, d - 2\}$;
	
	\noindent (c) $(-1)^{d - 2} \psi^{(d - 2)}$ is decreasing and convex in the interval $(a, b)$.
	
	%Moreover, if $\psi$ has derivatives of any order in $(a, b)$ and if $(-1)^k \psi^{(k)}(w) \geqslant 0$ for all $w \in (a, b)$ and for all $k \in \mathbb{Z}_+$ (the set of the non-negative integers), then $\psi$ is also said to be completely monotone.
\end{definition}
The following result yields a characterization for the Archimedean copulas $C_\psi$.
\begin{theorem}[\citealt*{DuSe2015}, p. 198]
	Let $\psi(w) : [0, \infty)$ be a generator and $d \geqslant 2$ a fixed integer. The following statements are equivalent:
	
	\noindent (a) $\psi$ is $d$-monotone on $[0, \infty)$;
	
	\noindent (b) The function $C_\psi : I^d \mapsto I$ is a $d$-dimensional copula.
\end{theorem}
In this work, five Archimedean copulas are addressed for the development of survival copula models, which are presented below through their multivariate representations for the copula function and the associated generator, following mainly \cite{HJoe2015} for the copula model expressions. All of them specify a unique parameter $\theta$ to model the copula dependence.

\noindent\textbf{\emph{Ali-Mikhail-Haq (AMH) Copula}}

The Ali-Mikhail-Haq (AMH) copula \citep{AlMH1978} was originally proposed as a bivariate extension for the univariate logistic distribution. Starting from the generator $\psi_\theta(w) = (1 - \theta)/(\exp(w) - \theta)$, where $\theta \in [0, 1]$, the general expression for the $d$-dimensional AMH copula is given by
\begin{equation}\label{eq3.15AMHcop}
C(\boldsymbol{u} | \theta) = \dfrac{1 - \theta}{\prod_{j=1}^{d} \left[\dfrac{1 - \theta(1 - u_j)}{u_j}\right] - \theta}, \quad \boldsymbol{u} \in I^d\\
\end{equation}
If $d=2$, expression \eqref{eq3.15AMHcop} can be extended to accommodate negative dependence (\citealt*{HJoe1997}; \citealt*{HJoe2015}). In that case, it reduces to $C(u_1, u_2) = u_1 u_2 / [1 - \theta(1 - u_1)(1 - u_2)]$, $\theta \in [-1, 1]$.

\noindent\textbf{\emph{Clayton Copula}}

The Clayton copula \citep{Clay1978} has born as a survival model to demonstrate association between pairs of subjects for a disease incidence. Its dependence parameter $\theta$ can assume any value in the interval $[-1/(d-1), \infty)$, with generator $\psi_\theta(w) = \left[\max(0, 1 + \theta w)\right]^{-1 / \theta}$. However, in the bivariate case (for instance), if $\theta \in [-1, 0)$, the copula is restricted to a region that satisfies $\psi_\theta^{-1}(u_1) + \psi_\theta^{-1}(u_2) < \psi_\theta^{-1}(0) = -1/\theta$ \citep{Coor2018}. Thus, it depends on $\theta$ and the density is $0$ on the set $\{(u_1, u_2): u_1^{-\theta} + u_2^{-\theta} < 1\}$ \citep{HJoe2015}. To avoid this, only the non-negative part of the support, $\theta \in [0, \infty)$, will be addressed here. In that case, the $d$-dimensional Clayton copula is expressed as
\begin{equation}\label{eq3.17Fracop}
C(\boldsymbol{u} | \theta) = \left\{\max\left[\sum_{j=1}^{d} u_j^{-\theta} - (d - 1), \, 0\right]\right\}^{-1/\theta}, \quad \boldsymbol{u} \in I^d.
\end{equation}

\noindent\textbf{\emph{Frank Copula}}

The Frank copula \citep{Fran1979} arose on a purely mathematical context, but with many properties already discovered in its debut. Starting from the generator $\psi_\theta(w) = -\log[1 - (1 - \exp^{-\theta}) \exp^{-w}] / \theta$, where $\theta \in (0, \infty)$, the general expression for the $d$-dimensional Frank copula is given by
\begin{equation}
C(\boldsymbol{u} | \theta) = -\dfrac{1}{\theta} \log\left\{1 - \dfrac{\prod_{j=1}^{d} \left[1 - \exp(-\theta u_j)\right]}{\left[1 - \exp(-\theta)\right]^{d - 1}}\right\}, \quad \boldsymbol{u} \in I^d.
\end{equation}
Like in the AMH copula, the expression in \eqref{eq3.17Fracop} can be extended to allow negative dependence if $d=2$, reducing to $C(u_1, u_2) = -(1/\theta)$ $\log\left\{\left[1 - \exp^{-\theta} - \left(1 - \exp^{-\theta u_1}\right)\left(1 - \exp^{-\theta u_2}\right)\right] / \left(1 - \exp^{-\theta}\right)\right\}$, $\theta \in \mathbb{R} \ \{0\}$.

\noindent\textbf{\emph{Gumbel-Hougaard (GH) Copula}}

The Gumbel-Hougaard (GH) copula \citep{Nels2006} was originally proposed by \cite{Gumb1960} and later discussed by \cite{Houg1986} in the survival analysis context. The generator for this copula model is given by $\psi_\theta(w) = \exp(-w^{1/\theta})$, where $\theta \in [1, \infty)$. The general expression for the $d$-dimensional GH copula is given by
\begin{equation}
C(\boldsymbol{u} | \theta) = \exp\left\{-\left[\sum_{j=1}^{d} \left(-\log(u_j)\right)^\theta\right]^{1 / \theta}\right\}, \quad \boldsymbol{u} \in I^d.
\end{equation}

\noindent\textbf{\emph{Joe Copula}}

The Joe copula first appeared in another work of \cite{Fran1981}, but its properties have been investigated in depth by \cite{HJoe1993}. %, initially specifying two parameters to model the copula dependence. For the purpose of this work, the specification considered is the one in Section 3.1 of the second paper, but taking only $\theta$ as the dependence parameter (the other being fixed at $1$). 
Taking the generator $\psi_\theta(w) = 1 - \left[1 - \exp\left(-w\right)\right]^{1/\theta}$, where $\theta \in [1, \infty)$, the general expression for the $d$-dimensional Joe copula is given by
\begin{equation}
C(\boldsymbol{u} | \theta) = 1 - \left\{1 - \prod_{j=1}^{d} \left[1 - \left(1 - u_j\right)^\theta\right]\right\}^{1 / \theta}, \quad \boldsymbol{u} \in I^d.
\end{equation}

\subsubsection{Measuring Dependence}

The most common way to summarize dependence (\emph{e.g.}) between two random variables $X_1$ and $X_2$ is through a measure of monotone association, \emph{i.e.}, as their relationship approaches a monotone function in probability, the measure should increase in absolute value. In the copula theory, a monotone measure of association is a function $q(\theta; X_1, X_2)$ that satisfies the following properties (see \citealt*{HJoe2015}, p. 54 for more details):
\begin{enumerate}
	\item \textbf{Domain}: $q(\theta; X_1, X_2)$ can be defined for all pairs of random variables;
	\item \textbf{Symmetry}: $q(\theta; X_1, X_2) = q(\theta; X_2, X_1)$;
	\item \textbf{Range}: $q(\theta; X_1, X_2) \in [-1, 1]$;
	\item \textbf{Independence}: If $X_1$ and $X_2$ are independent, then $q(\theta; X_1, X_2) = 0$;
	\item \textbf{Invariance}: If $h_1, h_2$ are strictly increasing functions, then $q[\theta; h_1(X_1), h_2(X_2)] = q(\theta; X_1, X_2)$.
\end{enumerate}

%As Hofert \emph{et al.} \citeyearpar{Hofe2018} pointed out, the (monotone) Pearson's  linear correlation does not attend all properties above: its domain and independence can only be evaluated if both $X_1$ and $X_2$ have finite second-order moments. Moreover, this measure is not invariant under strictly increasing nonlinear transformations.
By only depending on a single dependence parameter, the Kendall's $\tau$ correlation measure satisfies the properties above.
\begin{definition}[\citealt*{HJoe2015}, p. 55]
	Let $(X_1, X_2)$ and $(X^*_1, X^*_2)$ be two independent random pairs with a common joint continuous distribution $F_{12}$ and copula $C$. The Kendall's $\tau$ measure is given in terms of $C$ by
	\begin{align}
	\tau % &= \mathbb{P}\left[\left(X_1 - X^*_1\right) \left(X_2 - X^*_2\right) > 0\right] - \mathbb{P}\left[\left(X_1 - X^*_1\right) \left(X_2 - X^*_2\right) < 0\right] \nonumber \\
	% &= 2 \mathbb{P}\left[\left(X_1 - X^*_1\right) \left(X_2 - X^*_2\right) > 0\right] - 1 \nonumber \\
	% &= 4 \int F_{12} dF_{12} - 1 \nonumber \\
	&= 4 \int_{[0,1]^2} C(u_1, u_2) dC(u_1, u_2) - 1. \label{eq3.17ktau} %\nonumber \\
	% &= 4 \left[\dfrac{1}{2} - \int_{[0,1]^2} C_{2|1}(u_2|u_1) C_{1|2}(u_1|u_2) du_1 du_2\right] - 1 \quad (\text{through integration by parts}) \nonumber \\
	% &= 1 - 4\left[\int_{[0,1]^2} C_{2|1}(u_2|u_1) C_{1|2}(u_1|u_2) du_1 du_2\right].
	\end{align}
\end{definition}
Since $\tau$ depends on a known copula with well-defined marginal conditional distributions, the coefficient can be directly estimated by plugging the estimate for $\theta$ from the survival copula modeling in \eqref{eq3.17ktau}. For Archimedean copulas, the Kendall's $\tau$ can be rewritten in terms of the corresponding generator as (\citealt*{Nels2006}, p. 163 and 166)
\begin{equation*}
\tau = 4 \int_0^1 \dfrac{\psi^{-1}(w)}{\psi^{-1;(1)}(w)} dw + 1 = 1 - 4 \int_0^{\infty} w \left[\dfrac{d \psi(w)}{dw}\right]^2 dw.
\end{equation*}
Table \ref{tab3.1} presents the Kendall's $\tau$ measure as function of $\theta$ for the AMH, Clayton, Frank, GH and Joe copulas. Computing of those coefficients will be done through functions of the R \texttt{copula} package Hofert \emph{et al.} \citeyearpar{copu2020}.
\begin{table}[htb]
	\caption{Kendall's $\tau$ coefficient for some bivariate Archimedean copula models \citep{HJoe2015}}
	\label{tab3.1}
	\centering
	\begin{tabular}{cccc}
		\toprule
		Copula & Kendall's $\tau$ & Copula & Kendall's $\tau$ \\
		\midrule
		AMH & $\dfrac{3 \theta - 2}{3 \theta} - \dfrac{2 (1 - \theta)^2 \log(1 - \theta)}{3 \theta^2}$ & GH & $\dfrac{\theta - 1}{\theta}$
		\vspace{1mm}\\
		\vspace{1mm}
		Clayton & $\dfrac{\theta}{\theta + 2}$ & Joe & $1 + \dfrac{2}{2 - \theta} \left[\Psi(2) - \Psi\left(\dfrac{2}{\theta} + 1\right)\right]$
		\vspace{1mm}\\
		\vspace{1mm}
		Frank & $1 + \dfrac{4}{\theta} \left[D_1(\theta) - 1\right]$ & &
		\\
		\bottomrule
	\end{tabular}
\end{table}

The special functions in Table \ref{tab3.1} are the Dèbye function $D_k(x) = k x^{-k} \int_0^x a^k \left[\exp(a) - 1\right]^{-1} da$, with $k \in \mathbb{N}$, and the digamma function $\Psi(x) = (d/dx) \log\left[\Gamma(x)\right]$.

\subsection{Baseline Distribution}

Provided the framework for copula modeling as a joint survival function, it is time to look for their marginal survival functions. In the context of independent right-censoring, a survival function is generally composed of two terms: a baseline distribution to model the hazard (or odds) function behavior given a reference level, and a regression class to model covariates (if there is any) through a function with positive image. %Any family of probabilistic models with support on the positive real line is a candidate to be a baseline distribution.
Following \cite{Coll2015}, let $T \geqslant 0$ be a non-negative random variable for the survival time of a subject and $t$ the value of its actual observed time. Suppose that $T$ has a baseline distribution with continuous density function $f(t | \boldsymbol{\kappa})$, where $\boldsymbol{\kappa}$ is the vector of parameters from the baseline distribution, %. Then, the cumulative distribution function of $T$ is given by
%\begin{equation}\label{eq1.1cdf}
%F(t | \boldsymbol{\kappa}) = P(T < t | \boldsymbol{\kappa}) = \int_{0}^{t} f(u | \boldsymbol{\kappa}) du,
%\end{equation}
%representing the probability that a survival time is less than a value $t$.
and cumulative distribution function $F(t | \boldsymbol{\kappa})$. The survival function is defined as the probability that a survival time is greater than or equal to a value $t$, %so
%\begin{equation}\label{eq1.2sf}
%S(t | \boldsymbol{\kappa}) = P(T \geqslant t | \boldsymbol{\kappa}) = 1 - F(t | \boldsymbol{\kappa})
%\end{equation}
%is the complement of $F(t | \boldsymbol{\kappa})$ and represents the probability that a subject will survive beyond a given time $t$.
\emph{i.e.}, $S(t | \boldsymbol{\kappa}) = P(T \geqslant t | \boldsymbol{\kappa}) = 1 - F(t | \boldsymbol{\kappa})$. In order to allow more meaningful interpretations, survival modeling generally do not deal directly with % \eqref{eq1.1cdf} or \eqref{eq1.2sf}
$F(t | \boldsymbol{\kappa})$ or $S(t | \boldsymbol{\kappa})$, but instead with the hazard or odds functions. %in their formulation.
The hazard function of a baseline distribution can be defined as
\begin{equation}\label{eq1.3hf}
h(t | \boldsymbol{\kappa}) = \dfrac{f(t | \boldsymbol{\kappa})}{S(t | \boldsymbol{\kappa})} = -\dfrac{d}{dt} \{\log[S(t | \boldsymbol{\kappa})]\}.
\end{equation}
Integrating \eqref{eq1.3hf} with respect to $t$, its cumulative $H(t | \boldsymbol{\kappa})$ is obtained. In its turn, the odds function is given by
\begin{equation}\label{eq1.4of}
R(t | \boldsymbol{\kappa}) = \dfrac{F(t | \boldsymbol{\kappa})}{S(t | \boldsymbol{\kappa})} = \dfrac{1 - \exp\left[-H(t | \boldsymbol{\kappa})\right]}{\exp\left[-H(t | \boldsymbol{\kappa})\right]} = \exp\left[H(t | \boldsymbol{\kappa})\right] - 1.
\end{equation}
Differentiating \eqref{eq1.4of} with respect to $t$, we have
\begin{equation}\label{eq1.5dof}
r(t | \boldsymbol{\kappa}) = \dfrac{d}{dt} \left[R(t | \boldsymbol{\kappa})\right] = %\dfrac{d}{dt} \left\{\exp\left[H(t | \boldsymbol{\kappa})\right] - 1\right\} =
h(t | \boldsymbol{\kappa})\exp\left[H(t | \boldsymbol{\kappa})\right] = \dfrac{f(t | \boldsymbol{\kappa})}{\left[S(t | \boldsymbol{\kappa})\right]^2}.
\end{equation}
Defined all the required expressions, %to model the hazard or odds function,
a family of probabilistic models must be chosen for the baseline distribution. However, this choice cannot be arbitrary: it should represent well many plausible possibilities for the empirical survival curve behavior. Having that in mind, three models for the baseline distribution are addressed here and presented below.

\noindent\textbf{\emph{Weibull Model}}

\cite{Weib1951} proposed a family of probability distributions for describing the life length of materials. A random variable $T$ follows a Weibull distribution with parameters $\boldsymbol{\kappa} = (\alpha, \lambda)$, $\lambda > 0,\,\alpha > 0$, if
\begin{gather*}
F(t | \alpha, \lambda) = 1 - \exp(-\lambda t^{\alpha}), \quad
f(t | \alpha, \lambda) = \lambda \alpha t^{\alpha - 1} \exp(-\lambda t^{\alpha}),
\end{gather*}
where $\lambda$ and $\alpha$ are the scale and shape parameters, respectively. Note that the classical exponential distribution is obtained if $\alpha = 1$. Regarding the hazard function and its cumulative, their expressions are given by
\begin{gather*}
h(t | \alpha, \lambda) = \lambda \alpha t^{\alpha - 1}, \quad
H(t | \alpha, \lambda) = \lambda t^{\alpha}.
\end{gather*}
Although simple, $h(t | \alpha, \lambda)$ accommodates increasing ($\alpha > 1$), decreasing ($\alpha < 1$) and constant ($\alpha = 1$, backing to the exponential case) behaviors for the hazard function of $t$. Even when increasing, $h(t | \alpha, \lambda)$ can be concave ($\alpha \in (1, 2)$), linear ($\alpha = 2$) or convex ($\alpha > 2$). However, it does not allow non-monotonicity, such as unimodal and ``bathtub'' forms. Nevertheless, the Weibull hazard function is a good start point for the development and comparison of any proposed survival model. Regarding the odds function and its derivative, they are expressed as
\begin{gather*}
R(t | \alpha, \lambda) = \exp(\lambda t^{\alpha}) - 1, \quad
r(t | \alpha, \lambda) = \lambda \alpha t^{\alpha - 1} \exp(\lambda t^{\alpha}).
\end{gather*}

\noindent\textbf{\emph{Bernstein Polynomial Model}}

The Bernstein Polynomials (BP) were originally proposed by \cite{Bern1913} as a proof for the Weierstrass Approximation Theorem in the unit interval \citep{Lore1986}. Compared to other polynomial approximations, the BP approximation has optimal shape-preserving property \citep{CaPe1993}. Chang \emph{et al.} \citeyearpar{Chan2005} noted that the finite BP approximation could be used to estimate both hazard and cumulative hazard functions. Assuming $t \in [0, \upsilon]$, where $\upsilon = \inf\{t: S(t)=0\} < \infty$, let $H(t)$ be the target function. Its BP approximation is given by
\begin{equation}\label{eq2.11BPchf}
BP_{(m)}(t; H) = \sum_{k=0}^{m} H\left(\dfrac{k}{m}\upsilon\right) b_{(k,m)} \left(\dfrac{t}{\upsilon}\right), \quad t \in [0, \upsilon],
\end{equation}
and its first derivative with respect to the time $t$, approximating the hazard function $h(t)$, is expressed as
\begin{equation}
\dfrac{d}{dt}BP_{(m)}(t; H) = BP_{(m)}(t; h) = \sum_{k=1}^{m} \left\{H \left(\dfrac{k}{m} \upsilon\right) - H \left(\dfrac{k-1}{m} \upsilon\right)\right\} \left(\dfrac{1}{\upsilon}\right) f_{B} \left(\dfrac{t}{\upsilon}; k, m-k+1\right), \label{eq2.12derBPchf}
\end{equation}
where $B$ denotes the Beta distribution $B(k, m - k + 1)$. For simplicity, the cumulative hazards differences between braces and the two last terms in \eqref{eq2.12derBPchf} will be rewritten, respectively, as
\begin{equation*}
\gamma_k = \left\{H \left(\dfrac{k}{m} \tau\right) - H \left(\dfrac{k-1}{m} \tau\right)\right\}, \quad
g_{(k, m)}(t) = \left(\dfrac{1}{\tau}\right) f_{B} \left(\dfrac{t}{\tau}; k, m-k+1\right).
\end{equation*}
Note that $\gamma_k > 0$, $k \in \{1, \ldots, m\}$, since $H(\cdot)$ is monotone increasing. As all $\gamma_k$ do not depend on $t$, no information is given on the true cumulative hazard function and all coefficients must be estimated, forming a vector $\boldsymbol{\kappa} = \boldsymbol{\gamma} = (\gamma_1, \ldots, \gamma_m)'$ of BP parameters. Given $t$, define also a vector $\boldsymbol{g}_{m}(t) = (g_{(1, m)}(t), \ldots, g_{(m, m)}(t))'$ of fixed non-negative quantities. Then, the hazard function and its cumulative are modeled as (\citealt*{OsGh2012}, p. 561)
\begin{gather*}
h(t | \boldsymbol{\gamma}) = \boldsymbol{\gamma}'\boldsymbol{g}_{m}(t), \quad
H(t | \boldsymbol{\gamma}) = \int_{0}^{t} h(u, \boldsymbol{\gamma}) du = \boldsymbol{\gamma}'\boldsymbol{G}_{m}(t),
\end{gather*}
where $\boldsymbol{G}_{m}(t) = (G_{(1, m)}(t), \ldots, G_{(m, m)}(t))'$, with
\begin{equation*}
G_{(k, m)}(t) = \int_{0}^{t} g_{(k, m)}(u) du = \int_{0}^{t} f_{B} \left(\dfrac{u}{\tau}; k, m-k+1\right) d\left(\dfrac{u}{\tau}\right) \geqslant 0, \quad k \in \{1, \ldots, m\}.
\end{equation*}

Alternatively, BP can approximate the odds function (and its derivative). Since the true cumulative hazard (or odds) function is unknown, a finite value should be taken for $m$ on the estimation of BP parameters. \cite{OsGh2012} suggest a value $m$ such that $n^{2/5} < m < n^{2/3}$ for the polynomial degree. As BP models are computationally intensive, this work will choose the smallest possible integer, $m = \lceil n^{2/5} \rceil$, for the simulation results and applied data in this work.

\noindent\textbf{\emph{Piecewise Exponential Model}}

Proposed by \cite{KaPr1973} as an alternative to the Cox's regression model in the presence of ties for the survival times or grouped survival data, the Piecewise Exponential (PE) model assumes that the hazard function is constant (\emph{i.e.}, an Exponential model) between consecutive distinct survival times. Then, the true hazard function is approximated by ``steps'' of constant hazard functions. The formal definition of a PE model starts from a finite partition of the time axis, \emph{i.e.}, a time grid $E = \{e_0, e_1, \ldots, e_p\}$, with $0 = e_0 < e_1 < \cdots < e_p < \infty$. That way, there are $p$ intervals $E_k = (e_{k-1}, e_k]$, $k = 1, \ldots, p$. For each interval, a constant hazard function is assumed, that is
\begin{equation*}%\label{eq2.14PEhaz}
h(t) = \lambda_k, \quad t \in E_k, \quad k = 1,\ldots,p.
\end{equation*}
Therefore, $\boldsymbol{\lambda} = (\lambda_1, \ldots, \lambda_p)$ is the vector of constant hazard rates. If $p=1$, the Exponential model is obtained as particular case. To obtain the cumulative hazard and odds functions, for $k = 1, \ldots, p$, define $t_k \in \{e_{k - 1}, t, e_k\}$ if $t \leqslant e_{k-1}$; $t \in E_k$ or $t > e_k$, respectively.
%\begin{equation*}%\label{eq2.15PEtim}
%t_k = \begin{cases}
%e_{k - 1}, & t \leqslant e_{k-1}; \\
%t          & t \in E_k;           \\
%e_k        & t > e_k.
%\end{cases}
%\end{equation*}
Then, the cumulative hazard function is %computed from \eqref{eq2.14PEhaz} as
expressed as
\begin{gather*}
H(t | \boldsymbol{\lambda}, E) = \sum_{k=1}^{p} \lambda_k(t_k - e_{k-1}) = \boldsymbol{\lambda}'\left(\boldsymbol{t} - \boldsymbol{e}\right).
\end{gather*}
Like BP models, the PE model accommodates a variety of shapes for the hazard function, %across the extension of $E$,
since the number of intervals $p$ is arbitrary and can be as large as needed. The choice of a fixed $p$ for the PE model has been widely investigated in the literature (see \cite{Silv2016} and references therein for a discussion). To retain comparability with the number $m$ of polynomial degrees from the BP model, this work will fix $p = \lceil n^{2/5} \rceil$ for the simulation results and applied data.

\subsection{The Yang-Prentice Model}

If there is information on a set of covariates for the subjects under study, a regression structure can be defined to model all covariates along with the baseline distribution for the hazard (or odds) function. Let $X$ be a design matrix represent all covariate information. Then, let $\boldsymbol{x}_i$ be the covariate values given a subject $i$ and $\boldsymbol{\beta}$ the parameters associated to each covariate. Then, the reference level is represented by a subject $i$ whose covariate values are all equal to zero ($\boldsymbol{x}_i = 0$).

To accommodate crossing survival curves, which cannot be dealt by PH and PO models, \cite{YaPr2005} proposed a new regression model class to situations where, \emph{e.g.}, a treatment can be effective in the long run but may present adverse effects in early stages of a follow-up. The Yang-Prentice (YP) model defines two vectors of short and long-term hazard ratio parameters to allow intersection between survival curves. Let $T \geqslant 0$ be a random variable for the survival time; $\boldsymbol{x}_i$ the vector of covariate values for a subject $i$, and $\Phi_i = (\phi_{i}^{(S)}, \phi_{i}^{(L)})$, with $\phi_{i}^{(S)} = \exp(\boldsymbol{x}_i \boldsymbol{\beta}^{(S)})$; $\phi_{i}^{(L)} = \exp(\boldsymbol{x}_i \boldsymbol{\beta}^{(L)})$, and $\boldsymbol{\beta}^{(S)}$, $\boldsymbol{\beta}^{(L)}$ are vectors of regression parameters with same length, neither of them including an intercept. The survival function for the YP model is given by \citep{Dema2019}
\begin{equation}\label{eq2.24ypsf}
S(t | \Phi_i) = \left[1 + \dfrac{\phi_{i}^{(S)}}{\phi_{i}^{(L)}} R_0(t)\right]^{-\phi_{i}^{(L)}},
\end{equation}
where $R_0(t) = \exp[H_0(t)] - 1$ is the baseline odds function. If $\boldsymbol{x}_i = \boldsymbol{0}$, then \eqref{eq2.24ypsf} reduces to the baseline survival function $S_0(t) = 1 / [1 + R_0(t)]$. The hazard function associated with \eqref{eq2.24ypsf} can be expressed as
\begin{equation}\label{eq2.25yphf}
h(t | \Phi_i) = \dfrac{\phi_{i}^{(S)} \phi_{i}^{(L)} r_0(t)}{\phi_{i}^{(S)} + \phi_{i}^{(L)} R_0(t)} = \dfrac{\phi_{i}^{(S)} \phi_{i}^{(L)}}{\phi_{i}^{(S)} F_0(t) + \phi_{i}^{(L)} S_0(t)} h_0(t).
\end{equation}
The YP model has some interesting properties to be highlighted. First, it can be seen from \eqref{eq2.24ypsf} and \eqref{eq2.25yphf} that both PH and PO models arise as particular cases by putting $\boldsymbol{\beta}^{(S)} = \boldsymbol{\beta}^{(L)}$ and $\boldsymbol{\beta}^{(L)} = \boldsymbol{0}$, respectively. Also, it can be shown that crossing survival curves are obtained if ${\beta}^{(S)}_l {\beta}^{(L)}_l < 0$ for any $l = 1, \ldots, q$, where $q$ is the number of covariates \citep{YaPr2005}. Finally, from \eqref{eq2.25yphf}, it follows that
\begin{equation*}
\lim\limits_{t \rightarrow 0} \dfrac{h(t | \Phi_i, \boldsymbol{x}_i)}{h(t | \Phi_i, \boldsymbol{0})} = \exp(\boldsymbol{x}_i \boldsymbol{\beta}^{(S)}) = \phi_{i}^{(S)}, \quad \lim\limits_{t \rightarrow \infty} \dfrac{h(t | \Phi_i, \boldsymbol{x}_i)}{h(t | \Phi_i, \boldsymbol{0})} = \exp(\boldsymbol{x}_i \boldsymbol{\beta}^{(L)}) = \phi_{i}^{(L)}.
\end{equation*}
Therefore, $\phi_{i}^{(S)}$ and $\phi_{i}^{(L)}$ can be interpreted as the short and long-term hazard ratios for a subject $i$, respectively, and $\boldsymbol{\beta}^{(S)}$, $\boldsymbol{\beta}^{(L)}$ are the correspondent vectors of short and long-term coefficients. Given a baseline function with a vector of parameters $\boldsymbol{\kappa}$, the survival likelihood function for the YP model can be expressed as
\begin{equation*}
L_{\text{YP}}(\boldsymbol{\kappa}, \boldsymbol{\beta}^{(S)}, \boldsymbol{\beta}^{(L)} | \boldsymbol{y}, X) \propto \prod_{i=1}^{n} \left[\dfrac{\phi_{i}^{(S)} \phi_{i}^{(L)}}{\phi_{i}^{(S)} F_0(y_i | \boldsymbol{\kappa}) + \phi_{i}^{(L)} S_0(y_i | \boldsymbol{\kappa})} h_0(y_i | \boldsymbol{\kappa})\right]^{\delta_i} \left[1 + \dfrac{\phi_{i}^{(S)}}{\phi_{i}^{(L)}} R_0(y_i | \boldsymbol{\kappa})\right]^{-\phi_{i}^{(L)}}.
\end{equation*}
In the presence of a covariate representing a treatment and control indicator variable for a subject, the YP model can provide continuous crossing survival functions given both values on that covariate. In this case, there exists a time point at which the survival curves intersect each other. Although the observed Fisher information matrix allows to obtain point and interval estimates for the YP model parameters, it is not straightforward to find an interval estimate for the crossing survival time $t^*$, since there is no closed form expression for the standard error of its estimator $\widehat{t^*}$ \citep{Dema2019}. A viable solution is to apply a numerical procedure to find the root that solves the equation $S_C(t^*) - S_T(t^*) = 0$, where $S_C(\cdot)$ and $S_T(\cdot)$ are the survival functions for control and treated subjects, respectively (leaving all other covariates constant), and then run a resampling method to enable inference for the crossing time.

%\subsection{Proposed Survival Copula Models}
\subsection{Proposed Survival Copula Modeling}

The previous subsections showed (i) the copula theory and its applications for univariate survival functions as components of the Archimedean copula class; (ii) baseline distributions, given a reference level, for the hazard or odds functions of (marginal) survival models, and (iii) a wide regression model class, the YP model, as the main structure to specify (marginal) covariate information. Therefore, survival copula models combining these three frameworks are presented and characterized by their corresponding survival copula likelihood function below. 

Let $n$ be the size of a random sample of clusters, where each cluster $i$ has $d$ marginal observed times $y_{i;1}, \ldots, y_{i;d}$ with their corresponding censoring indicator values $\delta_{i;1}, \ldots, \delta_{i;d}$, $j = 1, \ldots, d$. Then, let $\boldsymbol{y} = (\boldsymbol{y}_1, \ldots, \boldsymbol{y}_d)$ be a sample of observed times, $d_i = \sum_{j=1}^{d} \delta_{i;j}$, and $X = (X_1, \ldots, X_d)$ be an array of design matrices. Define a collection $\mathcal{P} = \{\theta, \boldsymbol{\kappa}, \boldsymbol{\beta}^{(S)}, \boldsymbol{\beta}^{(L)}\}$ of parameters for a copula $C_\psi$, a baseline distribution and a YP regression structure, respectively, with $\boldsymbol{\kappa} = (\boldsymbol{\kappa}_1, \ldots, \boldsymbol{\kappa}_d)$ being parameters from the same family of baseline distributions. %The following survival (Archimedean) copula models are proposed.
Then, we propose the following general expression for the likelihood function of a survival Archimedean copula model by replacing all the marginal survival and density functions in \eqref{eq3.6copllcontrib} for baseline and YP regression terms, given by
\begin{align}
L_{\overline{C}} (\boldsymbol{y} | X, \theta, \boldsymbol{\kappa}, \boldsymbol{\beta}^{(S)}, \boldsymbol{\beta}^{(L)})
&=\mathlarger{\prod_{i=1}^n} \dfrac{\partial^{d_i}}{\partial \prod_{j=1}^d \left(y_{i;j}\right)^{\delta_{i;j}}} \psi_\theta \left\{\psi_\theta^{-1} \left[\sum_{j=1}^d \left(1 + \dfrac{\phi_{i}^{(S)}}{\phi_{i}^{(L)}} R_0(y_{i;j}) \right)^{-\phi_{i}^{(L)}}\right]\right\} \nonumber \\
&\times \prod_{j=1}^d \left\{ \dfrac{\phi_{i}^{(S)} \phi_{i}^{(L)} h_0(y_{i;j})}{\phi_{i}^{(S)} F_0(y_{i;j}) + \phi_{i}^{(L)} S_0(y_{i;j})} \left[1 + \dfrac{\phi_{i}^{(S)}}{\phi_{i}^{(L)}} R_0(y_{i;j}) \right]^{-\phi_{i}^{(L)}} \right\}^{\delta_{i;j}},
\label{eq4.1}
\end{align}
where $\phi_{i}^{(S)} = \exp(\boldsymbol{x}_i \boldsymbol{\beta}^{(S)})$ and $\phi_{i}^{(L)} = \exp(\boldsymbol{x}_i \boldsymbol{\beta}^{(L)})$. Thereby, maximum likelihood estimates for all parameters are obtained by maximizing directly the logarithm for expression \eqref{eq4.1} through %numerical maximization routines. % If your R package is on CRAN and published in JSS, cite it here!
%To fulfill this task, programmed functions in Stan (\citeyear{Stan2020}) for
the Broyden-Fletcher-Goldfarb-Shanno (BFGS) algorithm \citep{NoWr2006}. %were used in this paper.
Standard errors of each parameter estimator are computed from the observed Fisher information matrix, obtained by inverting the approximated log-likelihood Hessian matrix. %provided by Stan.
In the section that follows, some asymptotic properties of the maximum likelihood estimates are empirically investigated through an extensive Monte Carlo (MC) simulation study.

\section{Numerical Results}

This section presents a MC simulation study to evaluate the performance of bivariate Archimedean survival copula models taking $M = 1000$ replications of data sets. To generate them, the R \texttt{copula} package \citep{copu2020} was used to obtain marginal uniform realizations $u_{i;j}$, $i = 1, \ldots, n$ and $j = 1, 2$, for the %five Archimedean copulas addressed in this paper (AMH, Clayton, Frank, GH and Joe).
AMH, Clayton, Frank, GH and Joe copulas. For each margin, the same design matrix was specified, with two covariates $X_{i, 1} \sim \text{Bern}(0.5)$ and $X_{i, 2} \sim \text{N}(0, 1)$ generated independently and identically distributed for all $i$. Assuming $u_{i;j} = S(t_{i;j})$, %where $S(\cdot)$ is a survival function with inverse $S^{-1}(\cdot)$,
each marginal survival time was generated as $t_{i;j} = S^{-1}(u_{i;j})$. Two baseline distributions were considered for generation: the traditional Weibull model, and the Exponentiated Weibull (EW) model proposed by \cite{MuSr1993}, which has high flexibility for the hazard function behavior, allowing unimodal and bathtub forms not accommodated by the Weibull model. For the EW model, $\boldsymbol{\kappa} = (\alpha, \lambda, \xi)$, where %$\alpha$ and $\lambda$ are the shape and scale parameters from the Weibull model, and
$\xi$ is the exponentiation parameter. %If $\xi = 1$, the Weibull model is obtained.
Each baseline distribution used for marginal generation %of marginal survival times
was combined with one of three regression model classes: PH, PO, or YP. %Thereby, each marginal survival time $t_{i;j}$ was generated from the following expressions in \eqref{eq5.1} for the YP class, obtaining the PH and PO as particular cases by taking $\boldsymbol{\beta}_j^{(S)} = \boldsymbol{\beta}_j^{(L)}$ or $\boldsymbol{\beta}_j^{(L)} = \boldsymbol{0}$ (respectively), and depending on the chosen baseline distribution (Weibull or EW).
% \begin{equation}\label{eq5.1}
% t_{i;j} = \begin{cases}
% \left\{\log\left[1 - q\left(u_{i;j}, \boldsymbol{x}_{i;j}, \boldsymbol{\beta}_j^{(S)}, \boldsymbol{\beta}_j^{(L)}\right)\right] \lambda_j^{-1}\right\}^{\left(\alpha_j^{-1}\right)}; \quad &\text{(Weibull YP)} \\ %Ok!
% \left\{-\log\left[1 - \left(1 - \left(1 - q\left(u_{i;j}, \boldsymbol{x}_{i;j}, \boldsymbol{\beta}_j^{(S)}, \boldsymbol{\beta}_j^{(L)}\right)\right)^{-1}\right)^{\xi_j^{-1}}\right]\lambda_j^{-1}\right\}^{\left(\alpha_j^{-1}\right)}, \quad &\text{(EW YP)} %Ok!
% \end{cases}
% \end{equation}
%with
% \begin{equation*}
% q\left(u_{i;j}, \boldsymbol{x}_{i;j}, \boldsymbol{\beta}_j^{(S)}, \boldsymbol{\beta}_j^{(L)}\right) = \dfrac{\phi_{i}^{(L)}}{\phi_{i}^{(S)}} \left(1 - u_{i;j}^{-1/\phi_{i}^{(L)}}\right).
% \end{equation*}
Given a survival copula model, both baseline and regression structures are generated from the same family for all margins. Simulation scenarios are primarily defined by a fixed sample size ($n = 500$) for all margins, and three Kendall's $\tau$ values ($\tau \in \{0.25, 0.5, 0.75\}$). %In other words, there are three scenarios for study with the same sample size: S1, with $\tau = 0.25$, S2, with $\tau = 0.50$, and S3, with $\tau = 0.75$.
These true values for $\tau$ were chosen in order to obtain results given distinct levels of dependence, and to retain comparability among all fitted copulas: for a same value of $\tau$, $\theta$ can be quite different from one copula to another. %The Kendall's $\tau$ choice as the measure of association for this work is justified by its easier computation if compared to Spearman's $\rho$ for all copulas (as showed in \ref{tab3.1}), and the recurrent use of Kendall's $\tau$ in the literature, as seen in \cite{Pren2017}.
%Although the baseline distribution and regression model class are from the same family for both margins, the chosen values for %$\boldsymbol{\kappa}_j$ and $\boldsymbol{\beta}_j$ (for PH or PO), or $\boldsymbol{\kappa}_j$, $\boldsymbol{\beta}_j^{(S)}$, and $\boldsymbol{\beta}_j^{(L)}$ (for YP)
%$\boldsymbol{\kappa}_j$, $\boldsymbol{\beta}_j^{(S)}$, and $\boldsymbol{\beta}_j^{(L)}$, $j = 1, 2$, are different. %Regardless of copula or regression model class,
For the Weibull model as the baseline generator, $\boldsymbol{\kappa}_1 = (\alpha_1, \lambda_1) = (1.2, 0.8)$ and $\boldsymbol{\kappa}_2 = (\alpha_2, \lambda_2) = (1.6, 1.2)$. Both parameter specifications yield increasing marginal hazard functions. If the generator process for the baseline is the EW model, $\boldsymbol{\kappa}_1 = (\alpha_1, \lambda_1, \xi_1) = (2.1, 0.5, 0.3)$, and $\boldsymbol{\kappa}_2 = (\alpha_2, \lambda_2, \xi_2) = (2.5, 0.6, 0.2)$. Those parameter specifications produce ``bathtub''-shaped marginal hazard functions. Concerning the regression model classes, %regardless of copula or baseline distribution, %$\boldsymbol{\beta}_1 = (-0.7, 0.4)$ and $\boldsymbol{\beta}_2 = (-0.9, 0.6)$ for PH and PO models. For the YP model,
$\boldsymbol{\beta}_1^{(S)} = (-0.7, 0.4)$, $\boldsymbol{\beta}_2^{(S)} = (-0.9, 0.6)$, $\boldsymbol{\beta}_1^{(L)} = (0.8, -0.6)$, and $\boldsymbol{\beta}_2^{(L)} = (1.0, -0.8)$.
To introduce censoring, all generated times $t_{i; j}$, $i = 1, \ldots, n$ and $j = 1, 2$, were compared to a threshold value $a_{i; j}$ sampled from a continuous uniform distribution $\text{U}(0, a_j)$. If $t_{i; j} \leqslant a_{i; j}$, then $y_{i; j} = t_{i; j}$ is a failure time. Otherwise, $y_{i; j} = a_{i; j}$ is a censored time. The values $a_j$ were chosen in order to achieve a failure rate between $65$\% and $85$\% for each margin. %, given their corresponding set of parameters values, for all scenarios.
The threshold choices were $a_1 = 6$, $a_2 = 4$ when generating from Weibull baseline, and $a_1 = 4$, $a_2 = 3$ from EW.

The simulation study has three main goals: (i) given the same baseline function and regression model class, to compare results when fitting all five Archimedean copulas to a set of data generated from a given copula, (ii) given the correct fitting for the copula and regression structure, to compare results for different fitted baseline functions, and (iii) for a same combination of copula and baseline, evaluate the fitting of nested regression model classes. %when generating from a given one.
Goal (i) is achieved by presenting results for regression parameter estimation and information criteria, while goal (ii) is reached by showing results for Kendall's $\tau$ estimation. To achieve goal (iii), an analysis through the Likelihood Ratio (LR) test is done %, confronting pairs of 
for pairs of nested models given each generated regression model class (supposed unknown for each test). % Since fitted baseline functions from different families (Weibull, BP or PE) have distinct specifications for $\boldsymbol{\kappa}_j$, $j = 1, 2$, which do not allow to compare them directly, their parameter estimation is not addressed here.
If the data is marginally generated from a Weibull baseline function, correctly fitting both copula and regression structures, fitted Weibull models are expected to have great performance. Also, semiparametric fitted models (BP and PE) are expected to perform well due to its nonparametric appeal. However, for marginal generation from an EW baseline function, semiparametric models are expected to perform better than the Weibull fitting due to their flexibility in capture non-monotonic behaviors potentially present in both hazard functions. To evaluate MC estimates for regression parameters %set $\boldsymbol{\beta}_j$ (for PH and PO), or $\boldsymbol{\beta}_j^{(S)}, \boldsymbol{\beta}_j^{(L)}$ (for YP)
and %the Kendall's $\tau$ correlation as function of the original dependence parameter 
$\tau$ as function of
$\theta$, the following MC statistics were computed. %over results for all $M$ data sets.
If $\nu$ is a parameter %of interest
for inference:
\begin{itemize}
	\item The Average Estimate (AE) of $\nu$ is given by the mean of point estimates $\widehat{\nu}_l, l = 1, \ldots, M$;%:
	%\begin{equation}
	%\text{AE}(\nu) = \dfrac{1}{M} \sum_{l=1}^{M} \widehat{\nu}_l;
	%\end{equation}
	\item The Standard Deviation Estimate (SDE) of $\nu$ is given by the standard deviation of point estimates $\widehat{\nu}_l$;%:
	%\begin{equation}
	%\text{SDE}(\nu) = \left\{\dfrac{1}{M - 1} \sum_{l=1}^{M} \left[\widehat{\nu}_l - AE\left(\nu\right)\right]^2\right\}^{1/2};
	%\end{equation}
	\item The Average Standard Error (ASE) of $\nu$ is given by the mean of standard error estimates $\text{se}(\widehat{\nu}_l)$;%:
	%\begin{equation}
	%\text{ASE}(\nu) = \dfrac{1}{M} \sum_{l=1}^{M} \text{se}\left(\widehat{\nu}_l\right);
	%\end{equation}
	\item The Average Relative Bias (ARB) of $\nu$, generally expressed by a percentage, is given by the mean of relative biases $(\widehat{\nu}_l - \nu/|\nu|$ computed over all estimates $\widehat{\nu}_l$, with respect to the true value $\nu$;%:
	%\begin{equation}
	%\text{ARB}(\nu) = 100 \times \dfrac{1}{M} \sum_{l=1}^M \dfrac{\left(\widehat{\nu}_l - \nu\right)}{|\nu|};
	%\end{equation}
	\item The Coverage Rate (CR) of $\nu$ is the proportion of $M$ data sets that provides a interval with a pre-specified confidence level (95\%) that contains the true value $\nu$.
\end{itemize}
%On a well-fitted model, the AE is expected to be near the true parameter value, the SDE and ASE are expected to have values close to each other, the ARB is expected to be around $0$, and the CR is expected to be near the pre-specified confidence level. If the ARB is close to $0$, underestimating the true standard deviation of a given parameter (ASE < SDE) leads to a CR lower than the confidence level. When overestimating the true standard deviation (ASE > SDE), the CR is greater than the confidence level.

Another way to compare fitted models is computing an information criteria based on the log-likelihood. %$\ell = \ell(\boldsymbol{\nu})$ for a set of $n$ observed values, given the $p$-dimensional vector $\widehat{\boldsymbol{\nu}}$ of estimates.
A model is preferable if it has the lowest criteria value. For a frequentist approach, a method frequently used is the Akaike Information Criteria (AIC, \citealt*{Akai1974}). Comparisons for fitted models here will be done through the mean AIC and, given a copula model used for generation, using also the proportion of choice for each of the five fitted copulas. Furthermore, given any pair of fitted survival copula models nested with respect to the regression model class, an analysis through the LR test is done to conclude if the augmented model %(the one with more regression parameters)
is significant. It is expected that a fitted PH (or PO) model will be chosen due to their parsimony over YP model for PH (PO) marginally generated data, but also that a fitted YP model perform significantly better than fitted PH or PO models for YP marginally generated data.

%Defined all statistics, criteria and tests to be evaluated, simulation results are produced for all scenarios.
Due to the high number of tables and figures, along with the detection of similar patterns, presented results here are limited to the scenario where $\tau=0.25$ when generating from the YP class (except for LR tests), while the others can be accessed through the link \url{https://wrmfstat.shinyapps.io/CopRegEst/} (mainly for regression parameter estimation) or seen in the supplementary material. For the regression parameter estimates, results are restricted to the dichotomous covariate in the 1st copula margin, and given generated data from the AMH model, since its Kendall's $\tau$ value varies in the interval $[(5 - 8\log(2))/3, 1/3] \approx [-0.1817, 0.3333]$ (the only one that does not cover the open unit interval). This way, AMH generated data used a value of $\tau$ truncated to the upper limit when necessary. The idea is to show that this change has little effect on regression parameter estimates from different fitted copula models, since the true $\tau$ value is ``unknown''. Thus, results for AMH generated data are comparable to those of other Archimedean copulas used for generation, which in general appoints to similar conclusions.

Results for fitted survival copula models over simulated data are presented in the following subsections, divided by the marginally generated baseline distribution (Weibull or EW). Each subsection contains MC estimates for regression parameters (including the average lower and upper bounds -- ALB and AUB -- for their estimated intervals), AIC for the fitted models, MC estimates for Kendall's $\tau$ correlation, and LR tests for nested regression model classes. Finally, an additional simulation study is done over a specific scenario ($\tau=0.25$), and only for copula generated data with Weibull YP margins, %change later when EW results are ready.
to estimate marginal crossing survival times given a combination of fitted (correct) copula (one of the five discussed in this paper), baseline distribution (one of Weibull, BP or PE), and the YP model class. %To handle the solving of nonlinear equations involving survival functions of treated and control subjects in R, the command \texttt{uniroot} is used to find the corresponding root for each marginal crossing survival time \citep{R2022}.

\subsection{Generated Copulas with Weibull Baseline}

Results are presented below for fitted survival copula models over generated copula data with marginal Weibull baseline distribution, associated to the YP class, and divided by MC estimates for regression parameters, AIC for the fitted models, MC estimates for Kendall's $\tau$ correlation, and LR tests for nested fitted models.

\noindent\textbf{\emph{Regression Parameter Estimates}}

The MC estimates on regression parameters for fitted survival copula models are showed from Tables \ref{tab5.1.3} to \ref{tab5.1.4}, divided by fitted baseline distribution for each regression parameter set  from the YP class ($\boldsymbol{\beta}_j^{(S)}$ and $\boldsymbol{\beta}_j^{(L)}$), on the 1st copula margin ($j=1$). For those results, comparisons are done among fitted models with different copulas.%, but always maintaining the same regression model class used for generation.
\begin{table}[htb]
	\small
	\caption{MC statistics for 1st margin short-term regression parameter estimates of fitted survival copula models over AMH Weibull YP generated data ($n = 500$; $\tau = 0.25$)}
	\label{tab5.1.3}
	\centering	
	%\begin{tabular}{ccrrrrrrr}
	\begin{tabular}{cccrrrrrrr}
		\toprule
		%\multicolumn{1}{c}{\multirow{2}[2]{*}{Parameter}} &
		%\multicolumn{1}{c}{\multirow{2}[2]{*}{Copula}} &
		%\multicolumn{7}{c}{Weibull YP Fitting} \\
		%\cmidrule{3-9}
		% && AE & SDE & ASE & ARB (\%) & ALB & AUB & CR (\%) \\
		Parameter & Copula & Fitting & AE & SDE & ASE & ARB (\%) & ALB & AUB & CR (\%) \\
		\midrule
		$\beta_{11}^{(S)}=-0.7$ & AMH & Weibull YP & 
		-0.7077 & 0.1542 & 0.1526 & -1.0935 & -1.0099 & -0.4054 & 96.3783 \\ 
		& Clayton &  &
		-0.6939 & 0.1544 & 0.1545 & 0.8667 & -0.9965 & -0.3914 & 96.0765 \\ 
		& Frank &  &
		-0.7137 & 0.1543 & 0.1542 & -1.9526 & -1.0161 & -0.4112 & 96.4895 \\ 
		& GH &  &
		-0.7387 & 0.1537 & 0.1550 & -5.5346 & -1.0400 & -0.4375 & 94.8949 \\ 
		& Joe &  &
		-0.7281 & 0.1543 & 0.1565 & -4.0175 & -1.0306 & -0.4256 & 94.7791 \\
		% \midrule
		% $\beta_{12}^{(S)}=0.4$
		% & AMH & 0.3973 & 0.0917 & 0.0965 & -0.6857 & 0.2176 & 0.5770 & 94.0644 \\ 
		% & Clayton & 0.3938 & 0.0911 & 0.0958 & -1.5552 & 0.2152 & 0.5724 & 94.1650 \\ 
		% & Frank & 0.3946 & 0.0921 & 0.0978 & -1.3604 & 0.2141 & 0.5750 & 93.9819 \\ 
		% & GH & 0.3900 & 0.0917 & 0.0996 & -2.5020 & 0.2102 & 0.5698 & 93.5936 \\ 
		% & Joe & 0.3891 & 0.0920 & 0.1008 & -2.7225 & 0.2088 & 0.5694 & 93.0723 \\
		%\midrule
		%\multicolumn{1}{c}{\multirow{2}[2]{*}{Parameter}} &
		%\multicolumn{1}{c}{\multirow{2}[2]{*}{Copula}} & \multicolumn{7}{c}{BP YP Fitting} \\
		%\cmidrule{3-9}
		% && AE & SDE & ASE & ARB (\%) & ALB & AUB & CR (\%) \\
		\midrule
		Parameter & Copula & Fitting & AE & SDE & ASE & ARB (\%) & ALB & AUB & CR (\%) \\
		\midrule
		$\beta_{11}^{(S)}=-0.7$ & AMH & BP YP &
		-0.7047 & 0.1610 & 0.1609 & -0.6751 & -1.0202 & -0.3893 & 94.6894 \\ 
		& Clayton &  &
		-0.7031 & 0.1612 & 0.1619 & -0.4359 & -1.0189 & -0.3872 & 95.1000 \\ 
		& Frank &  &
		-0.7006 & 0.1616 & 0.1632 & -0.0923 & -1.0175 & -0.3838 & 94.9000 \\ 
		& GH &  &
		-0.7000 & 0.1618 & 0.1656 & 0.0033 & -1.0170 & -0.3830 & 95.0853 \\ 
		& Joe &  &
		-0.6980 & 0.1619 & 0.1673 & 0.2905 & -1.0152 & -0.3807 & 94.5892 \\
		% \midrule
		% $\beta_{12}^{(S)}=0.4$
		% & AMH & 0.3983 & 0.0920 & 0.0969 & -0.4133 & 0.2180 & 0.5787 & 94.0882 \\ 
		% & Clayton & 0.3965 & 0.0919 & 0.0968 & -0.8667 & 0.2165 & 0.5766 & 93.6000 \\ 
		% & Frank & 0.3956 & 0.0923 & 0.0976 & -1.0976 & 0.2147 & 0.5765 & 94.4000 \\ 
		% & GH & 0.3929 & 0.0916 & 0.0980 & -1.7673 & 0.2134 & 0.5724 & 93.6810 \\ 
		% & Joe & 0.3920 & 0.0918 & 0.1001 & -1.9878 & 0.2120 & 0.5721 & 93.0862 \\
		%\midrule
		%\multicolumn{1}{c}{\multirow{2}[2]{*}{Parameter}} &
		%\multicolumn{1}{c}{\multirow{2}[2]{*}{Copula}} & \multicolumn{7}{c}{PE YP Fitting} \\
		%\cmidrule{3-9}
		% && AE & SDE & ASE & ARB (\%) & ALB & AUB & CR (\%) \\
		\midrule
		Parameter & Copula & Fitting & AE & SDE & ASE & ARB (\%) & ALB & AUB & CR (\%) \\
		\midrule
		$\beta_{11}^{(S)}=-0.7$
		& AMH & PE YP &
		-0.7142 & 0.1604 & 0.1594 & -2.0219 & -1.0285 & -0.3998 & 95.0902 \\ 
		& Clayton &  &
		-0.7110 & 0.1604 & 0.1604 & -1.5749 & -1.0254 & -0.3966 & 95.1807 \\ 
		& Frank &  &
		-0.7121 & 0.1610 & 0.1615 & -1.7298 & -1.0277 & -0.3965 & 95.0853 \\ 
		& GH &  &
		-0.7101 & 0.1607 & 0.1640 & -1.4432 & -1.0251 & -0.3951 & 95.0853 \\ 
		& Joe &  &
		-0.7071 & 0.1606 & 0.1652 & -1.0206 & -1.0219 & -0.3924 & 94.8847 \\
		%\midrule
		% $\beta_{12}^{(S)}=0.4$
		% & AMH & 0.3899 & 0.0914 & 0.0951 & -2.5355 & 0.2107 & 0.5690 & 94.4890 \\ 
		% & Clayton & 0.3882 & 0.0912 & 0.0951 & -2.9483 & 0.2095 & 0.5670 & 94.0763 \\ 
		% & Frank & 0.3876 & 0.0917 & 0.0961 & -3.0900 & 0.2080 & 0.5673 & 94.5838 \\ 
		% & GH & 0.3878 & 0.0908 & 0.0969 & -3.0581 & 0.2097 & 0.5658 & 93.8816 \\ 
		% & Joe & 0.3869 & 0.0909 & 0.0985 & -3.2749 & 0.2087 & 0.5651 & 92.9789 \\
		\bottomrule
	\end{tabular}
\end{table}

For results in Table \ref{tab5.1.3}, %when correctly fitting the YP class,
the ARB for short-term parameters is always lower than 6\%, even when fitting the wrong copula, and lower than 3\% when choosing the correct one. In its turn, the CR is at most 0.03 away from the confidence level (set as 95\%) for all regression parameters, even when fitting the wrong copula. Correctly fitted (AMH) copula models had, in general, smaller ARB values and closer CR values to the confidence level. As expected, fitted semiparametric models (BP and PE) perform similar to (correctly) fitted Weibull models, but without imposing any parametric restriction for the (marginal) hazard rate function to obtain good regression parameter estimates.
\begin{table}[htb]
	\small
	\caption{MC statistics for 1st margin long-term regression parameter estimates of fitted survival copula models over AMH Weibull YP generated data ($n = 500$; $\tau = 0.25$).}
	\label{tab5.1.4}
	\centering	
	%\begin{tabular}{ccrrrrrrr}
	\begin{tabular}{cccrrrrrrr}
		\toprule
		%\multicolumn{1}{c}{\multirow{2}[2]{*}{Parameter}} &
		%\multicolumn{1}{c}{\multirow{2}[2]{*}{Copula}} &
		%\multicolumn{7}{c}{Weibull YP Fitting} \\
		%\cmidrule{3-9}
		% && AE & SDE & ASE & ARB (\%) & ALB & AUB & CR (\%) \\
		Parameter & Copula & Fitting & AE & SDE & ASE & ARB (\%) & ALB & AUB & CR (\%) \\
		\midrule
		$\beta_{11}^{(L)}=0.8$
		& AMH & Weibull YP & 0.8454 & 0.3206 & 0.3317 & 5.6743 & 0.2171 & 1.4737 & 95.8753 \\ 
		& Clayton &  & 0.8033 & 0.2984 & 0.3197 & 0.4179 & 0.2186 & 1.3881 & 93.7626 \\ 
		& Frank &  & 0.8734 & 0.3346 & 0.3453 & 9.1696 & 0.2175 & 1.5292 & 97.1916 \\ 
		& GH &  & 0.9268 & 0.3543 & 0.3693 & 15.8551 & 0.2325 & 1.6212 & 97.7978 \\ 
		& Joe &  & 0.9124 & 0.3655 & 0.3767 & 14.0457 & 0.1961 & 1.6287 & 97.9920 \\
		% \midrule
		% $\beta_{12}^{(L)}=-0.6$
		% & AMH & -0.5985 & 0.1231 & 0.1199 & 0.2430 & -0.8397 & -0.3574 & 95.1710 \\ 
		% & Clayton & -0.5832 & 0.1190 & 0.1197 & 2.8063 & -0.8164 & -0.3499 & 94.2656 \\ 
		% & Frank & -0.6063 & 0.1263 & 0.1218 & -1.0567 & -0.8538 & -0.3588 & 95.9880 \\ 
		% & GH & -0.6041 & 0.1280 & 0.1224 & -0.6776 & -0.8549 & -0.3532 & 96.3964 \\ 
		% & Joe & -0.6041 & 0.1288 & 0.1239 & -0.6877 & -0.8565 & -0.3517 & 96.4859 \\
		% \midrule
		%\multicolumn{1}{c}{\multirow{2}[2]{*}{Parameter}} &
		%\multicolumn{1}{c}{\multirow{2}[2]{*}{Copula}} & \multicolumn{7}{c}{BP YP Fitting} \\
		%\cmidrule{3-9}
		% && AE & SDE & ASE & ARB (\%) & ALB & AUB & CR (\%) \\
		\midrule
		Parameter & Copula & Fitting & AE & SDE & ASE & ARB (\%) & ALB & AUB & CR (\%) \\
		\midrule
		$\beta_{11}^{(L)}=0.8$
		& AMH & BP YP & 0.8732 & 0.3456 & 0.5894 & 9.1463 & 0.1756 & 1.5304 & 95.3815 \\ 
		& Clayton &  & 0.8440 & 0.3284 & 0.5070 & 5.5004 & 0.1889 & 1.4762 & 94.7948 \\ 
		& Frank &  & 0.8636 & 0.3519 & 0.4083 & 7.9500 & 0.1683 & 1.5479 & 96.2963 \\ 
		& GH &  & 0.8558 & 0.3507 & 0.4249 & 6.9739 & 0.1617 & 1.5363 & 96.0843 \\ 
		& Joe &  & 0.8777 & 0.3634 & 0.6558 & 9.7078 & 0.1415 & 1.5660 & 96.2814 \\
		% \midrule
		% $\beta_{12}^{(L)}=-0.6$
		% & AMH & -0.6073 & 0.1276 & 0.1256 & -1.2173 & -0.8574 & -0.3573 & 95.5868 \\ 
		% & Clayton & -0.5933 & 0.1248 & 0.1237 & 1.1244 & -0.8379 & -0.3485 & 95.4955 \\ 
		% & Frank & -0.6111 & 0.1302 & 0.1281 & -1.8558 & -0.8664 & -0.3559 & 95.9000 \\ 
		% & GH & -0.5996 & 0.1305 & 0.1288 & 0.0696 & -0.8554 & -0.3438 & 96.0883 \\ 
		% & Joe & -0.6027 & 0.1317 & 0.1301 & -0.4453 & -0.8607 & -0.3444 & 96.4895 \\
		%\midrule
		%\multicolumn{1}{c}{\multirow{2}[2]{*}{Parameter}} &
		%\multicolumn{1}{c}{\multirow{2}[2]{*}{Copula}} & \multicolumn{7}{c}{PE YP Fitting} \\
		%\cmidrule{3-9}
		% && AE & SDE & ASE & ARB (\%) & ALB & AUB & CR (\%) \\
		\midrule
		Parameter & Copula & Fitting & AE & SDE & ASE & ARB (\%) & ALB & AUB & CR (\%) \\
		\midrule
		$\beta_{11}^{(L)}=0.8$
		& AMH & PE YP & 0.8664 & 0.3452 & 0.3615 & 8.2977 & 0.1899 & 1.5429 & 95.9920 \\ 
		& Clayton &  & 0.8452 & 0.3268 & 0.3504 & 5.6472 & 0.2046 & 1.4858 & 95.7831 \\ 
		& Frank &  & 0.8727 & 0.3556 & 0.3698 & 9.0874 & 0.1757 & 1.5697 & 96.5898 \\ 
		& GH &  & 0.8601 & 0.3515 & 0.3650 & 7.5125 & 0.1711 & 1.5491 & 96.2889 \\ 
		& Joe &  & 0.8637 & 0.3570 & 0.3707 & 7.9609 & 0.1641 & 1.5633 & 96.4895 \\
		% \midrule
		% $\beta_{12}^{(L)}=-0.6$
		% & AMH & -0.5917 & 0.1249 & 0.1204 & 1.3806 & -0.8366 & -0.3469 & 96.0922 \\ 
		% & Clayton & -0.5805 & 0.1218 & 0.1191 & 3.2541 & -0.8192 & -0.3417 & 94.9799 \\ 
		% & Frank & -0.5958 & 0.1275 & 0.1226 & 0.6950 & -0.8458 & -0.3459 & 96.4895 \\ 
		% & GH & -0.5869 & 0.1276 & 0.1234 & 2.1866 & -0.8370 & -0.3368 & 96.4895 \\ 
		% & Joe & -0.5902 & 0.1290 & 0.1249 & 1.6344 & -0.8430 & -0.3374 & 96.1886 \\
		\bottomrule
	\end{tabular}
\end{table}

Again for correctly fitted YP models, but now looking results on long-term parameters in Table \ref{tab5.1.4}, the ARB is always below 10\% when fitting the correct copula, %being higher for the first (dichotomous) covariate than for the second (continuous) one,
regardless of the fitted baseline distribution. The CR values, however, are still near the confidence level, at most 0.03 away from it. The SDE and ASE estimates are also greater than those obtained for short-term parameters. %, specially for the dichotomous covariate.
These evidences implies more accurate estimates for short-term parameters than for their long-term counterpart. This is not a surprise: it is harder to estimate %$\boldsymbol{\beta}_j^{(L)}$ than $\boldsymbol{\beta}_j^{(S)}$
long-term parameters since there are fewer subjects under risk the closer a follow-up is to its end.

Even simulating with a high sample size, given generated data from an Archimedean copula model with marginal Weibull baseline distribution, fitted models with an incorrect (Archimedean) copula, among the five treated in this work (AMH, Clayton, Frank, GH and Joe), still show results close to the correct copula for regression parameters (regarding the ARB and CR) under a PH, PO or YP regression structure. Therefore, the choice of a copula for fitting has little impact on regression parameter estimation. Also, given a correct copula and regression fitting, ARB and CR values for fitted models with a nonparametric baseline (BP or PE) are similar to the obtained for (correctly) fitted Weibull models.

\noindent\textbf{\emph{Akaike Information Criteria}}

Unlike conclusions obtained for regression parameter tables, results for the mean of the AIC values and their proportion of choice (by the smallest AIC, given each MC replica) appoint clearly to the correct copula model choice, when generating marginally from the Weibull baseline distribution, as seen from Table \ref{tab5.2.3}. Looking only the correct copula fitted models, the highest proportions of choice are observed for the Joe copula, while the AMH and GH copulas (this last when fitted with the PE baseline) exhibit the lowest proportions, regardless of regression model class.
\begin{table}[htb]
	\small
	\caption{Mean AIC and choice proportion for all fitted survival copula models over generated data from each copula with Weibull YP margins ($n=500$; $\tau=0.25$).}
	\label{tab5.2.3}
	\centering	
	\begin{tabular}{ccrrrrrr}
		\toprule
		\multicolumn{1}{c}{\multirow{2}[2]{*}{True}} &
		\multicolumn{1}{c}{\multirow{2}[2]{*}{Fitted}} &
		\multicolumn{2}{c}{Weibull YP} &
		\multicolumn{2}{c}{BP YP} &
		\multicolumn{2}{c}{PE YP} \\
		\cmidrule{3-8}
		& & AIC & Choice (\%) & AIC & Choice (\%) & AIC & Choice (\%) \\
		\midrule
		AMH
		& AMH & 1452.21 & 70.10 & 1483.53 & 70.50 & 1484.45 & 68.90 \\ 
		& Clayton & 1456.01 & 17.50 & 1487.01 & 16.80 & 1487.78 & 18.00 \\ 
		& Frank & 1457.82 & 12.30 & 1489.12 & 12.50 & 1489.85 & 12.70 \\ 
		& GH & 1477.82 & 0.10 & 1506.95 & 0.20 & 1504.21 & 0.40 \\ 
		& Joe & 1492.98 & 0.00 & 1522.31 & 0.00 & 1519.60 & 0.00 \\
		\midrule
		Clayton
		& Clayton & 1443.29 & 83.10 & 1474.76 & 82.70 & 1475.45 & 83.60 \\ 
		& AMH & 1446.29 & 16.70 & 1477.91 & 16.90 & 1478.38 & 16.10 \\ 
		& Frank & 1461.58 & 0.20 & 1492.95 & 0.40 & 1493.41 & 0.30 \\ 
		& GH & 1481.45 & 0.00 & 1510.66 & 0.00 & 1507.64 & 0.00 \\ 
		& Joe & 1498.01 & 0.00 & 1527.91 & 0.00 & 1525.19 & 0.00 \\
		\midrule
		Frank
		& Frank & 1454.63 & 84.90 & 1486.48 & 83.10 & 1487.11 & 79.20 \\ 
		& AMH & 1459.53 & 12.30 & 1491.15 & 12.80 & 1491.76 & 12.40 \\ 
		& Clayton & 1469.78 & 0.50 & 1500.81 & 0.50 & 1501.42 & 0.40 \\ 
		& GH & 1469.61 & 2.30 & 1499.11 & 3.60 & 1496.36 & 8.00 \\ 
		& Joe & 1482.36 & 0.00 & 1511.19 & 0.00 & 1507.76 & 0.00 \\
		\midrule
		GH
		& GH & 1426.64 & 83.10 & 1458.16 & 79.00 & 1459.69 & 65.00 \\ 
		& AMH & 1460.84 & 0.10 & 1492.32 & 0.10 & 1493.69 & 0.00 \\ 
		& Clayton & 1472.77 & 0.00 & 1503.61 & 0.00 & 1504.59 & 0.00 \\ 
		& Frank & 1448.32 & 1.40 & 1479.62 & 1.60 & 1481.05 & 0.40 \\ 
		& Joe & 1430.85 & 15.40 & 1461.67 & 19.30 & 1461.39 & 34.60 \\
		\midrule
		Joe
		& Joe & 1392.68 & 90.00 & 1424.21 & 89.50 & 1425.65 & 95.40 \\
		& AMH & 1462.87 & 0.00 & 1494.04 & 0.00 & 1494.98 & 0.00 \\ 
		& Clayton & 1479.56 & 0.00 & 1510.41 & 0.10 & 1510.97 & 0.00 \\ 
		& Frank & 1441.53 & 0.00 & 1472.54 & 0.00 & 1473.74 & 0.00 \\ 
		& GH & 1398.89 & 10.00 & 1430.24 & 10.40 & 1434.29 & 4.60 \\ 
		\bottomrule
	\end{tabular}
\end{table}

For results in Table \ref{tab5.2.3} for correctly fitted YP models, and taking also the correct copula fitting, the Joe copula presents the smallest mean AIC values, given each fitted baseline. Inside each (correct) copula, fitted Weibull models exhibit the smallest mean AIC values, as expected. Fitted semiparametric models perform well and similarly to each other, with BP baseline are slightly better than PE for the majority of copulas, only being outperformed when given the Clayton copula.
Although the highest proportions of choice %by the smallest AIC
always point out to the correct copula, %choice, given any Archimedean copula for generation,
a similar pattern of fitted copulas with non-negligible proportions of choice can be identified, regardless of fitted baseline or regression structures. For example, when fitting %generated data from an AMH copula,
AMH generated data, fitted models with Clayton and Frank copulas always present proportions of choice higher than 10\%. The same can be said for fitted AMH models over data generated from Clayton or Frank copulas, and so on. Therefore, it is possible to define two groups of copulas that, albeit not nested on themselves with respect to the copula function, seem to capture similar behaviors of dependence (to be checked by correlation estimation below): the first one composed by Frank, AMH and Clayton copulas, and the second one by GH and Joe copulas.

\noindent\textbf{\emph{Correlation Estimates}}

As well as observed AIC results, the correct copula choice provides the best MC Kendall's $\tau$ estimates. %For the sake of simplicity, comments on the dependence parameter are restricted to the ARB results, since computations for a Kendall's $\tau$ standard error estimate from its analogue for $\theta$ (either through a Delta Method or a nonparametric resampling technique) is extremely difficult for some copulas or computationally demanding.
A special note should be taken on AMH models: as seen earlier, its Kendall's $\tau$ is restricted to the interval $[-0.1817, 0.3333]$. To accommodate a stronger (positive) dependence from AMH copula generated data, the supposed value for $\tau$ is truncated to the upper limit. That way, all other fitted survival copula models approximate well the truncated correlation value, but not the supposed original $\tau$ value according to the copula chosen for fitting (treating the true copula as unknown). This explains the higher negative values for the ARB of fitted models for AMH generated data with $\tau=0.5$ and $\tau=0.75$ (see the supplementary material for more details).
\begin{table}[thb]
	\small
	\caption{MC statistics for Kendall's tau estimates of fitted survival copula models over generated data from each copula model with Weibull YP margins ($n=500$; $\tau=0.25$)}
	\label{tab5.3.3}
	\centering	
	\begin{tabular}{ccrrrrrr}
		\toprule
		\multicolumn{1}{c}{\multirow{2}[2]{*}{True}} &
		\multicolumn{1}{c}{\multirow{2}[2]{*}{Copula}} &
		\multicolumn{2}{c}{Weibull YP} &
		\multicolumn{2}{c}{BP YP} &
		\multicolumn{2}{c}{PE YP} \\
		\cmidrule{3-8}
		& & AE & ARB (\%) & AE & ARB (\%) & AE & ARB (\%) \\
		\midrule
		AMH
		& AMH & 0.2504 & 0.1433 & 0.2506 & 0.2271 & 0.2513 & 0.5034 \\ 
		& Clayton & 0.2312 & -7.5121 & 0.2338 & -6.4906 & 0.2353 & -5.8715 \\ 
		& Frank & 0.2415 & -3.3931 & 0.2431 & -2.7794 & 0.2451 & -1.9716 \\ 
		& GH & 0.1798 & -28.0668 & 0.1956 & -21.7589 & 0.2091 & -16.3777 \\ 
		& Joe & 0.1114 & -55.4558 & 0.1286 & -48.5525 & 0.1454 & -41.8507 \\
		\midrule
		Clayton
		& Clayton & 0.2517 & 0.6780 & 0.2522 & 0.8913 & 0.2536 & 1.4278 \\ 
		& AMH & 0.2855 & 14.1892 & 0.2859 & 14.3573 & 0.2830 & 13.1897 \\ 
		& Frank & 0.2362 & -5.5211 & 0.2386 & -4.5782 & 0.2405 & -3.7861 \\ 
		& GH & 0.1742 & -30.3022 & 0.1916 & -23.3753 & 0.2054 & -17.8321 \\ 
		& Joe & 0.0974 & -61.0555 & 0.1124 & -55.0309 & 0.1301 & -47.9483 \\
		\midrule
		Frank
		& Frank & 0.2507 & 0.2944 & 0.2512 & 0.4942 & 0.2530 & 1.1973 \\ 
		& AMH & 0.2226 & -10.9621 & 0.2232 & -10.7273 & 0.2238 & -10.4709 \\ 
		& Clayton & 0.2100 & -15.9833 & 0.2150 & -13.9823 & 0.2164 & -13.4214 \\ 
		& GH & 0.1987 & -20.5014 & 0.2108 & -15.6613 & 0.2238 & -10.4984 \\ 
		& Joe & 0.1427 & -42.9139 & 0.1610 & -35.5982 & 0.1757 & -29.7065 \\
		\midrule
		GH
		& GH & 0.2488 & -0.4624 & 0.2511 & 0.4276 & 0.2663 & 6.5289 \\ 
		& AMH & 0.2106 & -15.7413 & 0.2117 & -15.3085 & 0.2121 & -15.1727 \\ 
		& Clayton & 0.2008 & -19.6602 & 0.2056 & -17.7519 & 0.2069 & -17.2430 \\ 
		& Frank & 0.2631 & 5.2348 & 0.2648 & 5.9315 & 0.2660 & 6.3892 \\ 
		& Joe & 0.2082 & -16.7336 & 0.2158 & -13.6746 & 0.2333 & -6.6855 \\
		\midrule
		Joe
		& Joe & 0.2516 & 0.6351 & 0.2546 & 1.8259 & 0.2732 & 9.2728 \\
		& AMH & 0.1968 & -21.2954 & 0.1979 & -20.8251 & 0.1983 & -20.6982 \\ 
		& Clayton & 0.1874 & -25.0559 & 0.1934 & -22.6403 & 0.1943 & -22.2730 \\ 
		& Frank & 0.2783 & 11.3159 & 0.2811 & 12.4523 & 0.2817 & 12.6789 \\ 
		& GH & 0.2803 & 12.1088 & 0.2784 & 11.3529 & 0.2945 & 17.8041 \\ 
		\bottomrule
	\end{tabular}
\end{table}

From Table \ref{tab5.3.3}, the least ARB was (in general) %almost always
observed for the correct copula choice. %(except for GH generated data when fitting the Frank copula with PE YP margins).
Moreover, given any fitted model with the correct copula, fitted Weibull models presented the lowest ARB among all fitted baselines (except for the GH copula) when correctly fitting the YP class. However, the BP baseline also performs well: their ARB values are always below 2\%. Concerning the PE baseline, it also performs well for the correct fitting of AMH, Clayton and Frank copulas (always below 2\%), but not as much as Weibull and BP models for GH and Joe copulas.

When simulating with a high sample size, even on a scenario with weak dependence, the choice of the correct Archimedean copula for fitting is crucial to ensure a suitable estimation of $\theta$ and consequently of the Kendall's $\tau$ correlation. %This was expected, since all copulas considered for fitting are defined as functions of a single dependence parameter, besides being enough to produce, on average, better AIC results.
Also, semiparametric models have similar performance to (correctly fitted) Weibull models for almost all fitted survival models with the correct copula, no matter which results are taken to comparison, be it regression parameters, an information criteria, or the correlation parameter. Concerning the identified groups of copulas earlier on mean AIC evaluation (Frank, AMH and Clayton composing the first group, and GH and Joe the second one), those are corroborated by ARB results on the correlation estimation.%, although the Frank copula also seems to compete with the correct fitting of a GH copula along with Joe.

\noindent\textbf{\emph{Likelihood Ratio Tests}}

The analysis presented below compares, through the Likelihood Ratio (LR) test, two nested models with respect to their regression structure (\emph{i.e.}, PH \emph{vs.} YP, and PO \emph{vs.} YP), given each class used for generation (always fitting the correct copula). %Note that there are no pairs of nested Archimedean copulas, or fitted baseline models, among those treated in this work.
Since there are two covariates for each margin, keeping the same specification for fitted PH, PO and YP models, p-values for all LR statistics from tests comparing nested models from each MC replica will be obtained from a $\chi^2$ distribution with $2 \times 2 = 4$ degrees of freedom (note that PH and PO classes have the same number of parameters). Under the null hypothesis, it is supposed that the additional regression parameters from YP model are not significant. If the LR statistic does not surpass the critical value under a significance of 5\% (here, equal to $\chi^2(0.95, 4) \approx 9.4877$), the regression class with less parameters (more parsimonious model), is chosen. Otherwise, the YP model is selected as the best class. Results for the average statistics and p-values from LR tests are presented from Tables \ref{tab5.4.1} to \ref{tab5.4.3}.
\begin{table}[htb]
	\small
	\caption{LR tests for PH and PO classes against YP class, given fitted survival copula models over generated data from each copula model with Weibull PH margins ($n=500; \tau=0.25$)}
	\label{tab5.4.1}
	\centering
	\begin{tabular}{ccrrrrrr}
		\toprule
		\multicolumn{1}{c}{\multirow{2}[2]{*}{Copula}} & \multicolumn{1}{c}{\multirow{2}[2]{*}{Class}} & \multicolumn{2}{c}{Weibull Fitting} & \multicolumn{2}{c}{BP Fitting} & \multicolumn{2}{c}{PE Fitting} \\
		\cmidrule{3-8}
		& & LR stat. & P-value & LR stat. & P-value & LR stat. & P-value \\
		\midrule
		AMH & PH & 4.1965 & 0.4846 & 4.2689 & 0.4787 & 4.1354 & 0.4907 \\ 
		& PO & 48.0752 & < 0.0001 & 33.8761 & 0.0005 & 34.1804 & 0.0002 \\ 
		\midrule
		Clayton & PH & 4.0989 & 0.4923 & 4.0398 & 0.4921 & 3.9649 & 0.4991 \\ 
		& PO & 48.7359 & < 0.0001 & 34.4732 & 0.0003 & 34.6441 & 0.0002 \\ 
		\midrule
		Frank & PH & 4.1500 & 0.4742 & 4.1365 & 0.4736 & 4.0266 & 0.4840 \\ 
		& PO & 47.6290 & < 0.0001 & 34.1908 & 0.0003 & 34.6151 & 0.0002 \\ 
		\midrule
		GH & PH & 4.0212 & 0.5008 & 4.0232 & 0.5000 & 3.8907 & 0.5124 \\ 
		& PO & 47.3462 & < 0.0001 & 33.5009 & 0.0003 & 34.0694 & 0.0002 \\ 
		\midrule
		Joe & PH & 4.1007 & 0.4867 & 4.1930 & 0.4755 & 4.0980 & 0.4858 \\ 
		& PO & 47.2089 & < 0.0001 & 33.6742 & 0.0004 & 34.7567 & 0.0003 \\
		\bottomrule
	\end{tabular}
\end{table}

For Table \ref{tab5.4.1}, when generating from the PH regression model class, all LR tests accept fitted YP models against the (incorrect) PO regression structure, given any combination of fitted baseline and copula, as expected. Remember that the YP model generalizes the PH class, which are not nested within the PO structure. However, the same cannot be said for LR tests confronting fitted PH and YP models. Their results are always non-significant, leading to the choice of (correctly) fitted PH models, since they are more parsimonious (fewer regression parameters). Therefore, given a large sample size, introducing more regression parameters with a wider functional form to capture both short and long-term covariate effects (without increasing the number of original covariates) does not provide a significantly better fitting.
\begin{table}[htb]
	\small
	\caption{LR tests for PH and PO classes against YP class, given fitted survival copula models over generated data from each copula model with Weibull PO margins ($n=500; \tau=0.25$)}
	\label{tab5.4.2}
	\centering
	\begin{tabular}{ccrrrrrr}
		\toprule
		\multicolumn{1}{c}{\multirow{2}[2]{*}{Copula}} & \multicolumn{1}{c}{\multirow{2}[2]{*}{Class}} & \multicolumn{2}{c}{Weibull Fitting} & \multicolumn{2}{c}{BP Fitting} & \multicolumn{2}{c}{PE Fitting} \\
		\cmidrule{3-8}
		& & LR stat. & P-value & LR stat. & P-value & LR stat. & P-value \\
		\midrule
		AMH & PH & 30.1272 & 0.0018 & 23.7872 & 0.0100 & 24.9197 & 0.0066 \\ 
		& PO & 5.5396 & 0.3600 & 4.1265 & 0.4960 & 5.3370 & 0.3465 \\ 
		\midrule
		Clayton & PH & 31.1509 & 0.0015 & 24.6702 & 0.0076 & 25.9595 & 0.0051 \\ 
		& PO & 5.5430 & 0.3617 & 4.0845 & 0.4890 & 5.2688 & 0.3469 \\ 
		\midrule
		Frank & PH & 29.9082 & 0.0017 & 23.6726 & 0.0096 & 24.7241 & 0.0066 \\ 
		& PO & 5.4930 & 0.3664 & 4.1064 & 0.4922 & 5.2970 & 0.3465 \\ 
		\midrule
		GH & PH & 29.0730 & 0.0026 & 23.2333 & 0.0096 & 24.0902 & 0.0063 \\ 
		& PO & 5.5658 & 0.3659 & 4.3068 & 0.4800 & 5.4802 & 0.3429 \\ 
		\midrule
		Joe & PH & 30.1505 & 0.0029 & 23.7688 & 0.0139 & 24.2719 & 0.0107 \\ 
		& PO & 5.5344 & 0.3654 & 4.2056 & 0.4780 & 5.4340 & 0.3316 \\
		\bottomrule
	\end{tabular}
\end{table}

Looking now for Table \ref{tab5.4.2}, this time generating from the PO class, the converse is also true: all LR tests accept fitted YP models against the (incorrect) PH structure, given any combination of fitted baseline and copula, as expected, since the YP model also generalizes the PO class. However, the same cannot be said for LR tests confronting fitted PO and YP models. Their results lead towards the choice of fitted PO models, since these are more parsimonious. Finally, for Table \ref{tab5.4.3}, when generating from the wider YP regression model class, all LR tests accept fitted YP models against the PH or PO structures, given any combination of fitted baseline and copula. This was also expected, since (nested) PH and PO models do not account for covariate short and long-term effects, but only during the whole time of follow-up.
\begin{table}[htb]
	\small
	\caption{LR tests for PH and PO classes against YP class, given fitted survival copula models over generated data from each copula model with Weibull YP margins ($n=500; \tau=0.25$)}
	\label{tab5.4.3}
	\centering
	\begin{tabular}{ccrrrrrr}
		\toprule
		\multicolumn{1}{c}{\multirow{2}[2]{*}{Copula}} & \multicolumn{1}{c}{\multirow{2}[2]{*}{Class}} & \multicolumn{2}{c}{Weibull Fitting} & \multicolumn{2}{c}{BP Fitting} & \multicolumn{2}{c}{PE Fitting} \\
		\cmidrule{3-8}
		& & LR stat. & P-value & LR stat. & P-value & LR stat. & P-value \\
		\midrule
		AMH & PH & 122.1934 & < 0.0001 & 108.0926 & < 0.0001 & 109.9563 & < 0.0001 \\ 
		& PO & 94.2805 & < 0.0001 & 80.5264 & < 0.0001 & 82.8430 & < 0.0001 \\ 
		\midrule
		Clayton & PH & 128.0063 & < 0.0001 & 113.2562 & < 0.0001 & 115.0452 & < 0.0001 \\ 
		& PO & 100.5764 & < 0.0001 & 85.5648 & < 0.0001 & 87.9450 & < 0.0001 \\ 
		\midrule
		Frank & PH & 122.1422 & < 0.0001 & 107.9013 & < 0.0001 & 109.4467 & < 0.0001 \\ 
		& PO & 96.4341 & < 0.0001 & 82.5658 & < 0.0001 & 84.6061 & < 0.0001 \\ 
		\midrule
		GH & PH & 120.1831 & < 0.0001 & 106.2706 & < 0.0001 & 107.3817 & < 0.0001 \\ 
		& PO & 93.6736 & < 0.0001 & 80.5333 & < 0.0001 & 82.1474 & < 0.0001 \\ 
		\midrule
		Joe & PH & 123.6797 & < 0.0001 & 108.7011 & < 0.0001 & 109.9063 & < 0.0001 \\ 
		& PO & 96.3288 & < 0.0001 & 82.1735 & < 0.0001 & 84.0249 & < 0.0001 \\
		\bottomrule
	\end{tabular}
\end{table}

Tables \ref{tab5.4.1} to \ref{tab5.4.3} showed that the analysis through LR tests for nested regression model classes, when generating from an Archimedean survival copula with marginal Weibull baseline distribution, %and regardless of the fitted copula or baseline models (among those considered in this work)
is a useful tool to choose the regression structure for fitting if the one that generated the data is unknown.

\subsection{Generated Copulas with EW Baseline}

Results for fitted survival copula models over generated data from Archimedean survival copulas with marginal EW baseline distribution, also associated to the YP class, are presented below, and divided in the same way as done for generated copula data with Weibull margins.

\noindent\textbf{\emph{Regression Parameter Estimates}}

The MC estimates on regression parameters for fitted survival copula models, when marginally generating from the EW distribution, are showed from Tables \ref{tab5.5.3} to \ref{tab5.5.4}, divided by fitted baseline distribution for each regression parameter set from the YP class ($j=1$). For those results, comparisons are again done among fitted models with different copulas.%, yet keeping the same regression model class used for generation.

% Reunite Weibull and EW results now?

\begin{table}[thb]
	\small
	\caption{MC statistics for 1st margin short-term regression parameter estimates of fitted survival copula models over AMH EW YP generated data ($n=500$; $\tau=0.25$)}
	\label{tab5.5.3}
	\centering	
	%\begin{tabular}{ccrrrrrrr}
	\begin{tabular}{cccrrrrrrr}
		\toprule
		%\multicolumn{1}{c}{\multirow{2}[2]{*}{Parameter}} &
		%\multicolumn{1}{c}{\multirow{2}[2]{*}{Copula}} &
		%\multicolumn{7}{c}{Weibull YP Fitting} \\
		%\cmidrule{3-9}
		% && AE & SDE & ASE & ARB (\%) & ALB & AUB & CR (\%) \\
		Parameter & Copula & Fitting & AE & SDE & ASE & ARB (\%) & ALB & AUB & CR (\%) \\
		\midrule
		$\beta_{11}^{(S)}=-0.7$
		& AMH & Weibull YP & -0.9417 & 0.1450 & 0.1420 & -34.5262 & -1.2259 & -0.6574 & 60.7692 \\ 
		& Clayton &  & -0.9342 & 0.1446 & 0.1425 & -33.4615 & -1.2177 & -0.6508 & 63.4615 \\ 
		& Frank &  & -0.9400 & 0.1457 & 0.1427 & -34.2802 & -1.2255 & -0.6545 & 61.5877 \\ 
		& GH &  & -0.9372 & 0.1473 & 0.1451 & -33.8843 & -1.2258 & -0.6486 & 63.3205 \\ 
		& Joe &  & -0.9184 & 0.1481 & 0.1474 & -31.2050 & -1.2086 & -0.6282 & 68.4547 \\
		% \midrule
		% $\beta_{12}^{(S)}=0.4$
		% & AMH & 0.3796 & 0.0941 & 0.1101 & -5.1048 & 0.1952 & 0.5640 & 90.1282 \\ 
		% & Clayton & 0.3837 & 0.0933 & 0.1078 & -4.0625 & 0.2009 & 0.5666 & 90.7692 \\ 
		% & Frank & 0.3685 & 0.0956 & 0.1131 & -7.8817 & 0.1811 & 0.5559 & 89.1165 \\ 
		% & GH & 0.3594 & 0.0960 & 0.1168 & -10.1546 & 0.1712 & 0.5476 & 86.3578 \\ 
		% & Joe & 0.3588 & 0.0967 & 0.1173 & -10.2903 & 0.1694 & 0.5483 & 86.5900 \\
		%\midrule
		%\multicolumn{1}{c}{\multirow{2}[2]{*}{Parameter}} &
		%\multicolumn{1}{c}{\multirow{2}[2]{*}{Copula}} & \multicolumn{7}{c}{BP YP Fitting} \\
		%\cmidrule{3-9}
		% && AE & SDE & ASE & ARB (\%) & ALB & AUB & CR (\%) \\
		\midrule
		Parameter & Copula & Fitting & AE & SDE & ASE & ARB (\%) & ALB & AUB & CR (\%) \\
		\midrule
		$\beta_{11}^{(S)}=-0.7$
		& AMH & BP YP & -0.7013 & 0.1591 & 0.1614 & -0.1835 & -1.0132 & -0.3894 & 95.1904 \\ 
		& Clayton &  & -0.7022 & 0.1595 & 0.1628 & -0.3190 & -1.0148 & -0.3896 & 94.8795 \\ 
		& Frank &  & -0.6912 & 0.1601 & 0.1643 & 1.2584 & -1.0049 & -0.3775 & 94.8000 \\ 
		& GH &  & -0.6949 & 0.1631 & 0.1683 & 0.7306 & -1.0146 & -0.3752 & 95.2381 \\ 
		& Joe &  & -0.6961 & 0.1633 & 0.1678 & 0.5590 & -1.0162 & -0.3760 & 95.6522 \\
		% \midrule
		% $\beta_{12}^{(S)}=0.4$
		% & AMH & 0.4284 & 0.0923 & 0.1003 & 7.0937 & 0.2475 & 0.6093 & 92.2846 \\ 
		% & Clayton & 0.4260 & 0.0921 & 0.1010 & 6.5109 & 0.2456 & 0.6065 & 92.1687 \\ 
		% & Frank & 0.4228 & 0.0926 & 0.1016 & 5.6939 & 0.2413 & 0.6043 & 92.5000 \\ 
		% & GH & 0.4168 & 0.0940 & 0.1037 & 4.2096 & 0.2326 & 0.6010 & 92.9215 \\ 
		% & Joe & 0.4138 & 0.0942 & 0.1038 & 3.4473 & 0.2292 & 0.5984 & 92.5831 \\
		%\midrule
		%\multicolumn{1}{c}{\multirow{2}[2]{*}{Parameter}} &
		%\multicolumn{1}{c}{\multirow{2}[2]{*}{Copula}} & \multicolumn{7}{c}{PE YP Fitting} \\
		%\cmidrule{3-9}
		% && AE & SDE & ASE & ARB (\%) & ALB & AUB & CR (\%) \\
		\midrule
		Parameter & Copula & Fitting & AE & SDE & ASE & ARB (\%) & ALB & AUB & CR (\%) \\
		\midrule
		$\beta_{11}^{(S)}=-0.7$
		& AMH & PE YP & -0.7154 & 0.1589 & 0.1576 & -2.1955 & -1.0268 & -0.4039 & 94.9648 \\ 
		& Clayton &  & -0.7125 & 0.1589 & 0.1584 & -1.7868 & -1.0240 & -0.4010 & 94.7844 \\ 
		& Frank &  & -0.7111 & 0.1596 & 0.1601 & -1.5865 & -1.0239 & -0.3983 & 94.8692 \\ 
		& GH &  & -0.7125 & 0.1611 & 0.1633 & -1.7907 & -1.0282 & -0.3968 & 95.5128 \\ 
		& Joe &  & -0.7111 & 0.1617 & 0.1639 & -1.5849 & -1.0280 & -0.3942 & 95.3668 \\
		% \midrule
		% $\beta_{12}^{(S)}=0.4$
		% & AMH & 0.4020 & 0.0913 & 0.0961 & 0.5080 & 0.2230 & 0.5810 & 93.9577 \\ 
		% & Clayton & 0.4002 & 0.0911 & 0.0960 & 0.0597 & 0.2218 & 0.5787 & 94.2828 \\ 
		% & Frank & 0.3988 & 0.0916 & 0.0968 & -0.3012 & 0.2193 & 0.5783 & 94.7686 \\ 
		% & GH & 0.3991 & 0.0921 & 0.0992 & -0.2233 & 0.2185 & 0.5797 & 93.0769 \\ 
		% & Joe & 0.3977 & 0.0927 & 0.0997 & -0.5820 & 0.2160 & 0.5794 & 92.9215 \\
		\bottomrule
	\end{tabular}
\end{table}

From Table \ref{tab5.5.3}, when marginally fitting a semiparametric (YP) model, the ARB for short-term parameters is always lower than 3\% for PE models and 8\% for BP models, even when fitting the wrong copula. However, for (incorrectly) fitted Weibull models, the ARB results are poor, being higher than 30\% and leading to CR values below 70\%. Therefore, for semiparametric models, fitting the correct copula has produced, in general, smaller ARB values and closer CR values to the confidence level, specially for the PE baseline. When looking the results for long-term parameters in Table \ref{tab5.5.4}, conclusions are similar to those obtained for short-term parameters in Table \ref{tab5.5.3}, but estimation for (incorrectly) fitted Weibull models is even poorer, with ARB values around or higher than 100\%. %, \emph{i.e.}, fitted semiparametric models performing well, but (incorrect) Weibull models presenting poor estimation.
%Along with results in Table \ref{tab5.5.3}, for the wider YP class,
Thus, fitted semiparametric models perform better than (incorrectly) fitted Weibull models, given copula generated data with marginal EW baseline distribution. As observed for marginally generated Weibull data, the SDE and ASE estimates are greater than those obtained for short-term parameters.%, specially for the dichotomous covariate. This is a consequence of the increased difficult on estimating long-term parameters due to the presence of fewer subjects under risk the closer a follow-up comes near its end.

\begin{table}[htb]
	\small
	\caption{MC statistics for 1st margin long-term regression parameter estimates of fitted survival copula models over AMH EW YP generated data ($n=500$; $\tau=0.25$)}
	\label{tab5.5.4}
	\centering
	%\begin{tabular}{ccrrrrrrr}
	\begin{tabular}{cccrrrrrrr}
		\toprule
		%\multicolumn{1}{c}{\multirow{2}[2]{*}{Parameter}} &
		%\multicolumn{1}{c}{\multirow{2}[2]{*}{Copula}} &
		%\multicolumn{7}{c}{Weibull YP Fitting} \\
		%\cmidrule{3-9}
		% && AE & SDE & ASE & ARB (\%) & ALB & AUB & CR (\%) \\
		Parameter & Copula & Fitting & AE & SDE & ASE & ARB (\%) & ALB & AUB & CR (\%) \\
		\midrule
		$\beta_{11}^{(L)}=0.8$
		& AMH & Weibull YP & 1.5848 & 1.4085 & 1.2081 & 98.0972 & -1.1837 & 4.3373 & 88.5751 \\ 
		& Clayton &  & 1.4957 & 0.5825 & 1.2353 & 86.9635 & 0.3539 & 2.6375 & 88.4615 \\ 
		& Frank &  & 1.7089 & 1.8974 & 1.5321 & 113.6081 & -2.0099 & 5.4277 & 90.3969 \\ 
		& GH &  & 1.8217 & 2.6840 & 1.7635 & 127.7085 & -3.4389 & 7.0822 & 89.1892 \\ 
		& Joe &  & 1.7977 & 2.6400 & 1.8409 & 124.7111 & -3.3923 & 6.9564 & 91.8159 \\
		% \midrule
		% $\beta_{12}^{(L)}=-0.6$
		% & AMH & -0.6478 & 0.1324 & 0.1224 & -7.9588 & -0.9073 & -0.3882 & 95.0000 \\ 
		% & Clayton & -0.6401 & 0.1286 & 0.1174 & -6.6816 & -0.8921 & -0.3881 & 95.6410 \\ 
		% & Frank & -0.6428 & 0.1393 & 0.1276 & -7.1338 & -0.9159 & -0.3698 & 95.7746 \\ 
		% & GH & -0.6269 & 0.1426 & 0.1329 & -4.4868 & -0.9063 & -0.3475 & 95.8816 \\ 
		% & Joe & -0.6221 & 0.1434 & 0.1348 & -3.6821 & -0.9031 & -0.3411 & 95.9132 \\
		%\midrule
		%\multicolumn{1}{c}{\multirow{2}[2]{*}{Parameter}} &
		%\multicolumn{1}{c}{\multirow{2}[2]{*}{Copula}} & \multicolumn{7}{c}{BP YP Fitting} \\
		%\cmidrule{3-9}
		% && AE & SDE & ASE & ARB (\%) & ALB & AUB & CR (\%) \\
		\midrule
		Parameter & Copula & Fitting & AE & SDE & ASE & ARB (\%) & ALB & AUB & CR (\%) \\
		\midrule
		$\beta_{11}^{(L)}=0.8$
		& AMH & BP YP & 0.8385 & 0.3274 & 0.4899 & 4.8121 & 0.1860 & 1.4694 & 95.1856 \\ 
		& Clayton &  & 0.8122 & 0.3103 & 0.4292 & 1.5309 & 0.1956 & 1.4120 & 94.0704 \\ 
		& Frank &  & 0.8570 & 0.3364 & 0.6558 & 7.1276 & 0.1859 & 1.5045 & 95.3954 \\ 
		& GH &  & 0.8562 & 0.4646 & 0.5860 & 7.0267 & -0.0543 & 1.7667 & 95.7529 \\ 
		& Joe &  & 0.8594 & 0.5793 & 0.5502 & 7.4200 & -0.2763 & 1.9946 & 95.6466 \\
		% \midrule
		% $\beta_{12}^{(L)}=-0.6$
		% & AMH & -0.6398 & 0.1257 & 0.1261 & -6.6401 & -0.8865 & -0.3938 & 93.6810 \\ 
		% & Clayton & -0.6227 & 0.1226 & 0.1232 & -3.7775 & -0.8629 & -0.3825 & 94.6787 \\ 
		% & Frank & -0.6432 & 0.1291 & 0.1295 & -7.1932 & -0.8961 & -0.3902 & 93.5000 \\ 
		% & GH & -0.6325 & 0.1304 & 0.1310 & -5.4141 & -0.8881 & -0.3769 & 94.0798 \\ 
		% & Joe & -0.6325 & 0.1307 & 0.1314 & -5.4090 & -0.8889 & -0.3766 & 94.1101 \\
		%\midrule
		%\multicolumn{1}{c}{\multirow{2}[2]{*}{Parameter}} &
		%\multicolumn{1}{c}{\multirow{2}[2]{*}{Copula}} & \multicolumn{7}{c}{PE YP Fitting} \\
		%\cmidrule{3-9}
		% && AE & SDE & ASE & ARB (\%) & ALB & AUB & CR (\%) \\
		\midrule
		Parameter & Copula & Fitting & AE & SDE & ASE & ARB (\%) & ALB & AUB & CR (\%) \\
		\midrule
		$\beta_{11}^{(L)}=0.8$
		& AMH & PE YP & 0.8352 & 0.3320 & 0.3398 & 4.4056 & 0.1845 & 1.4860 & 96.5760 \\ 
		& Clayton &  & 0.8138 & 0.3130 & 0.3264 & 1.7189 & 0.2003 & 1.4272 & 95.7874 \\ 
		& Frank &  & 0.8446 & 0.3448 & 0.3545 & 5.5735 & 0.1687 & 1.5204 & 97.1831 \\ 
		& GH &  & 0.8426 & 0.3412 & 0.3376 & 5.3285 & 0.1740 & 1.5113 & 97.0513 \\ 
		& Joe &  & 0.8501 & 0.3460 & 0.3431 & 6.2629 & 0.1720 & 1.5283 & 97.0399 \\
		% \midrule
		% $\beta_{12}^{(L)}=-0.6$
		% & AMH & -0.5922 & 0.1223 & 0.1172 & 1.2967 & -0.8319 & -0.3526 & 96.3746 \\ 
		% & Clayton & -0.5795 & 0.1188 & 0.1153 & 3.4173 & -0.8124 & -0.3466 & 95.5868 \\ 
		% & Frank & -0.5955 & 0.1253 & 0.1195 & 0.7452 & -0.8411 & -0.3500 & 96.3783 \\ 
		% & GH & -0.5908 & 0.1261 & 0.1225 & 1.5272 & -0.8381 & -0.3436 & 95.6410 \\ 
		% & Joe & -0.5946 & 0.1273 & 0.1230 & 0.8981 & -0.8440 & -0.3452 & 95.7529 \\
		\bottomrule
	\end{tabular}
\end{table}

Given generated data from an Archimedean copula model with marginal EW baseline distribution, again fitted models with an incorrect copula show results close to the correct one for regression parameters, regarding the ARB and CR, when marginally fitting a semiparametric baseline distribution. Also in this case, the choice of a copula for fitting has little impact on estimation of regression parameters. However, choosing a wrong parametric specification for the baseline distribution can lead to poor estimation on regression parameters. This was expected, since the hazard function for the Weibull model cannot accommodate non-monotone forms that can arise from marginally generated EW models.

\noindent\textbf{\emph{Akaike Information Criteria}}

For copula generated data with EW YP margins, results on the AIC of fitted BP and PE models corroborate to choose the correct copula, as seen from Table \ref{tab5.6.3}, except when generating from the Joe copula. In that case, fitted Frank and GH copulas have non-negligible or even higher proportions of choice, while fitted GH copula models also present slightly smaller mean AIC values, given the YP class. When taking other fitted models with the correct copula, the highest proportions of choice are observed for the Clayton and Frank copulas. On the other hand, for fitted models with the (incorrect) Weibull baseline, AIC results induce to the choice of Clayton copula instead of AMH when generating from the last one, although leading to the correct copula fitting for Clayton, Frank or GH generated data.

\begin{table}[htb]
	\small
	\caption{Mean AIC and choice proportion for all fitted survival copula models over generated data from each copula with EW YP margins ($n=500$; $\tau=0.25$)}
	\label{tab5.6.3}
	\centering	
	\begin{tabular}{ccrrrrrr}
		\toprule
		\multicolumn{1}{c}{\multirow{2}[2]{*}{True}} &
		\multicolumn{1}{c}{\multirow{2}[2]{*}{Fitted}} &
		\multicolumn{2}{c}{Weibull YP} &
		\multicolumn{2}{c}{BP YP} &
		\multicolumn{2}{c}{PE YP} \\
		\cmidrule{3-8}
		& & AIC & Choice (\%) & AIC & Choice (\%) & AIC & Choice (\%) \\
		\midrule
		AMH
		& AMH & 747.29 & 25.22 & 760.94 & 71.10 & 720.23 & 70.50 \\ 
		& Clayton & 745.76 & 73.63 & 764.42 & 19.50 & 723.36 & 18.50 \\ 
		& Frank & 757.80 & 1.15 & 767.78 & 9.40 & 726.24 & 11.00 \\ 
		& GH & 787.88 & 0.00 & 806.15 & 0.00 & 762.19 & 0.00 \\ 
		& Joe & 800.04 & 0.00 & 817.86 & 0.00 & 777.21 & 0.00 \\
		\midrule
		Clayton
		& Clayton & 738.31 & 98.70 & 749.94 & 85.40 & 709.60 & 85.60 \\ 
		& AMH & 742.98 & 1.30 & 753.99 & 14.40 & 713.19 & 14.20 \\ 
		& Frank & 761.62 & 0.00 & 770.33 & 0.20 & 729.10 & 0.20 \\ 
		& GH & 790.49 & 0.00 & 808.30 & 0.00 & 763.97 & 0.00 \\ 
		& Joe & 802.17 & 0.00 & 819.70 & 0.00 & 778.96 & 0.00 \\
		\midrule
		Frank
		& Frank & 746.74 & 57.25 & 762.06 & 83.50 & 721.45 & 85.50 \\ 
		& AMH & 747.55 & 31.15 & 766.39 & 15.50 & 726.34 & 13.70 \\ 
		& Clayton & 751.49 & 11.10 & 776.96 & 0.60 & 735.91 & 0.50 \\ 
		& GH & 771.96 & 0.50 & 793.08 & 0.40 & 750.36 & 0.30 \\ 
		& Joe & 785.56 & 0.00 & 805.68 & 0.00 & 764.70 & 0.00 \\
		\midrule
		GH
		& GH & 717.57 & 91.40 & 735.00 & 71.60 & 698.76 & 70.20 \\ 
		& AMH & 748.38 & 0.13 & 768.72 & 0.20 & 729.69 & 0.10 \\ 
		& Clayton & 754.07 & 0.38 & 781.10 & 0.00 & 741.32 & 0.00 \\ 
		& Frank & 737.14 & 7.08 & 755.41 & 28.10 & 716.00 & 28.70 \\ 
		& Joe & 726.33 & 1.01 & 744.60 & 0.10 & 707.78 & 1.00 \\
		\midrule
		Joe
		& Joe & 678.42 & 31.23 & 698.14 & 18.10 & 661.12 & 35.50 \\
		& AMH & 745.59 & 0.00 & 766.69 & 0.00 & 725.67 & 0.20 \\ 
		& Clayton & 756.43 & 0.00 & 784.82 & 0.00 & 742.78 & 0.00 \\ 
		& Frank & 723.87 & 0.13 & 743.68 & 22.50 & 702.14 & 22.20 \\ 
		& GH & 676.54 & 68.64 & 695.33 & 59.40 & 660.71 & 42.10 \\ 
		\bottomrule
	\end{tabular}
\end{table}

Although the majority of highest proportions of choice by the smallest AIC point out, in general, to the correct copula choice, it is possible to identify the same pattern (verified earlier for generated data with Weibull baseline) of fitted copulas with non-negligible proportions of choice seen earlier, regardless of fitted baseline or regression structure. Therefore, the same two copula groups (one involving Frank, AMH and Clayton, and the other composed by GH and Joe) that seem to capture similar behaviors of dependence are once again defined.

\noindent\textbf{\emph{Correlation Estimates}}

As well as observed AIC results for generated data with EW YP margins, the correct copula choice yields, in general, the best MC Kendall's $\tau$ estimates, as seen from Table \ref{tab5.7.3}, except when generating from GH or Joe copulas. For those two cases, fitted models with the correct copula present non-negligible (and negative) ARB values, %(always underestimating the true correlation value),
even when fitting a nonparametric baseline. This possibly evidences a difficulty on identifying the dependence parameter over more general behaviors for the marginal hazard function, with respect to some copulas.

\begin{table}[htb]
	\small
	\caption{MC statistics for Kendall's tau estimates of fitted survival copula models over generated data from each copula model with EW YP margins ($n=500$; $\tau=0.25$)}
	\label{tab5.7.3}
	\centering
	\begin{tabular}{ccrrrrrr}
		\toprule
		\multicolumn{1}{c}{\multirow{2}[2]{*}{True}} &
		\multicolumn{1}{c}{\multirow{2}[2]{*}{Copula}} &
		\multicolumn{2}{c}{Weibull YP} &
		\multicolumn{2}{c}{BP YP} &
		\multicolumn{2}{c}{PE YP} \\
		\cmidrule{3-8}
		& & AE & ARB (\%) & AE & ARB (\%) & AE & ARB (\%) \\
		\midrule
		AMH
		& AMH & 0.2681 & 7.2207 & 0.2480 & -0.8038 & 0.2502 & 0.0607 \\ 
		& Clayton & 0.2606 & 4.2588 & 0.2310 & -7.6188 & 0.2356 & -5.7568 \\ 
		& Frank & 0.2235 & -10.6162 & 0.2322 & -7.1067 & 0.2410 & -3.6106 \\ 
		& GH & 0.1209 & -51.6409 & 0.1021 & -59.1694 & 0.1394 & -44.2551 \\ 
		& Joe & 0.0596 & -76.1624 & 0.0446 & -82.1799 & 0.0669 & -73.2306 \\
		\midrule
		Clayton
		& Clayton & 0.2672 & 6.8785 & 0.2476 & -0.9431 & 0.2521 & 0.8494 \\ 
		& AMH & 0.2980 & 19.2168 & 0.2813 & 12.5057 & 0.2815 & 12.5824 \\ 
		& Frank & 0.2141 & -14.3685 & 0.2266 & -9.3537 & 0.2359 & -5.6588 \\ 
		& GH & 0.1099 & -56.0272 & 0.0943 & -62.2957 & 0.1326 & -46.9619 \\ 
		& Joe & 0.0484 & -80.6363 & 0.0354 & -85.8384 & 0.0545 & -78.2067 \\
		\midrule
		Frank
		& Frank & 0.2402 & -3.9398 & 0.2416 & -3.3708 & 0.2489 & -0.4216 \\ 
		& AMH & 0.2395 & -4.2144 & 0.2212 & -11.5299 & 0.2230 & -10.8022 \\ 
		& Clayton & 0.2525 & 1.0081 & 0.2134 & -14.6327 & 0.2174 & -13.0395 \\ 
		& GH & 0.1458 & -41.6675 & 0.1226 & -50.9588 & 0.1572 & -37.1380 \\ 
		& Joe & 0.0865 & -65.3876 & 0.0675 & -73.0175 & 0.0946 & -62.1548 \\
		\midrule
		GH
		& GH & 0.2006 & -19.7675 & 0.1800 & -27.9929 & 0.2125 & -15.0124 \\ 
		& AMH & 0.2267 & -9.3075 & 0.2098 & -16.0970 & 0.2112 & -15.5357 \\ 
		& Clayton & 0.2481 & -0.7549 & 0.2044 & -18.2425 & 0.2079 & -16.8314 \\ 
		& Frank & 0.2571 & 2.8342 & 0.2552 & 2.0885 & 0.2627 & 5.0931 \\ 
		& Joe & 0.1477 & -40.9160 & 0.1274 & -49.0377 & 0.1618 & -35.2800 \\
		\midrule
		Joe
		& Joe & 0.1899 & -24.0296 & 0.1673 & -33.0780 & 0.2061 & -17.5799 \\
		& AMH & 0.2148 & -14.0708 & 0.1970 & -21.1825 & 0.1977 & -20.9172 \\ 
		& Clayton & 0.2452 & -1.9213 & 0.1938 & -22.4956 & 0.1961 & -21.5458 \\ 
		& Frank & 0.2780 & 11.2183 & 0.2720 & 8.8198 & 0.2794 & 11.7567 \\ 
		& GH & 0.2370 & -5.1906 & 0.2150 & -13.9848 & 0.2468 & -1.2794 \\ 
		\bottomrule
	\end{tabular}
\end{table}

Given any fitted model with the correct copula, among AMH, Clayton, and Frank, the lowest ARB values were observed when fitting the PE model as the baseline distribution (always below 1\% in magnitude), followed by BP and Weibull models. %, regardless of the regression structure. 
Although not presenting good ARB values for correctly fitted models with GH or Joe copulas (above 10\%), the PE baseline still performs better than (incorrect) Weibull and BP models. %(in that order).
Comparing only the semiparametric models for those cases, using the BP baseline produces an ARB about twice the obtained when using the PE model. %(a bit more under the PH regression structure, and less under PO or YP classes).

From Table \ref{tab5.7.3}, it is possible to conclude that the choice of a correct Archimedean copula is necessary to ensure an appropriate estimation of $\tau$ if marginal survival times were generated by a more general process, but it is not sufficient depending on the true copula function. However, when looking to mean AIC values and proportions of choice for correctly fitted GH and Joe copulas from table \ref{tab5.6.3} it can be said that, if $\tau$ estimates alone are not the best, those models are better fitted if compared to almost all wrong copula choices, regardless of the fitted baseline distribution.

Finally, fitted PE models performed exceptionally better than the (also semiparametric) BP models and the (incorrect) Weibull models given any correct copula fitting, over generated data with EW margins. This contrasts with results obtained when generating marginally with the Weibull baseline, for which BP models had better fitting than PE ones for the majority of %copula and regression model class combinations
copulas (although still being outperformed to fitted Weibull models in general). That said, there is no immediate response for which semiparametric model is better when the generator process of copula data and marginal baseline distribution are unknown, but both BP and PE have proven to be useful.

\noindent\textbf{\emph{Likelihood Ratio Tests}}

For copula generated data with EW margins, the analysis  through LR tests %to be presented below
will not account for fitted Weibull models: for the purpose of this work, evaluation of nested models is done with respect to the regression model class. Therefore, %this evaluation
it is restricted to fitted semiparametric models. %, for which the total of baseline parameters that can be chosen is variable (fitted Weibull models will always estimate only two baseline parameters).
Again, p-values for all LR statistics will be obtained from a $\chi^2_{(4)}$. %distribution with $2 \times 2 = 4$ degrees of freedom.
Tables \ref{tab5.8.1} to \ref{tab5.8.3} present results for average statistics and p-values from LR tests on fitted BP and PE models over generated data with the EW baseline (under the correct copula fitting).

On Table \ref{tab5.8.1}, when generating from the PH class, all LR tests accept fitted YP models against the (incorrect) PO class, %, for any combination of fitted nonparametric baseline and copula,
as expected. The same cannot be said for LR tests confronting fitted PH and YP models, whose results are always non-significant, leading to the choice of (correctly and more parsimonious) fitted PH models. Thus, even on a more general baseline distribution for marginal survival times and given a large sample size, introducing more regression parameters to capture short and long-term covariate effects %(using the same original covariates)
does not provide a significantly better fitting.

\begin{table}[htb]
	\caption{LR tests for PH and PO classes against YP class, given fitted survival copula models over generated data from each copula model with EW PH margins ($n=500; \tau=0.25$)}
	\label{tab5.8.1}
	\centering
	%\begin{tabular}{ccrrrrrr}
	\begin{tabular}{ccrrrr}
		\toprule
		\multicolumn{1}{c}{\multirow{2}[2]{*}{Copula}} & \multicolumn{1}{c}{\multirow{2}[2]{*}{Class}} & %\multicolumn{2}{c}{Weibull Fitting} & 
		\multicolumn{2}{c}{BP Fitting} & \multicolumn{2}{c}{PE Fitting} \\
		%\cmidrule{3-8}
		\cmidrule{3-6}
		& % & LR stat. & P-value
		& LR stat. & P-value & LR stat. & P-value \\
		\midrule
		AMH & PH & 6.1175 & 0.3425 & 4.2940 & 0.4717 \\
		& PO & 26.0845 & 0.0172 & 30.8931 & 0.0009 \\
		\midrule
		Clayton & PH & 5.6297 & 0.3771 & 4.0687 & 0.4891 \\
		& PO & 26.4718 & 0.0124 & 31.3756 & 0.0008 \\
		\midrule
		Frank & PH & 5.9444 & 0.3527 & 4.1757 & 0.4726 \\
		& PO & 25.9596 & 0.0192 & 31.2596 & 0.0007 \\
		\midrule
		GH & PH & 5.4472 & 0.3895 & 4.1775 & 0.4853 \\
		& PO & 24.9161 & 0.0176 & 29.9006 & 0.0013 \\
		\midrule
		Joe & PH & 5.6756 & 0.3807 & 4.4143 & 0.4622 \\
		& PO & 23.1901 & 0.0283 & 28.8543 & 0.0023 \\
		\bottomrule
	\end{tabular}
\end{table}

Looking now for Table \ref{tab5.8.2}, this time generating from the PO class, the converse is also true: all LR tests accept fitted YP models against the (incorrect) PH regression structure, given any combination of fitted nonparametric baseline and copula. However, the same cannot be said for LR tests confronting fitted PO and YP models. Their results lead towards the choice of PO class, since its associated models are more parsimonious.

\begin{table}[htb]
	\caption{LR tests for PH and PO classes against YP class, given fitted survival copula models over generated data from each copula model with EW PO margins ($n=500; \tau=0.25$)}
	\label{tab5.8.2}
	\centering
	%\begin{tabular}{ccrrrrrr}
	\begin{tabular}{ccrrrr}
		\toprule
		\multicolumn{1}{c}{\multirow{2}[2]{*}{Copula}} & \multicolumn{1}{c}{\multirow{2}[2]{*}{Class}} & %\multicolumn{2}{c}{Weibull Fitting} & 
		\multicolumn{2}{c}{BP Fitting} & \multicolumn{2}{c}{PE Fitting} \\
		%\cmidrule{3-8}
		\cmidrule{3-6}
		& % & LR stat. & P-value
		& LR stat. & P-value & LR stat. & P-value \\
		\midrule
		AMH & PH & 27.9217 & 0.0071 & 26.7120 & 0.0051 \\
		& PO & 1.0745 & 0.7665 & 4.4347 & 0.4495 \\
		\midrule
		Clayton & PH & 28.9169 & 0.0040 & 27.9565 & 0.0038 \\
		& PO & 0.9686 & 0.7644 & 4.4032 & 0.4478 \\
		\midrule
		Frank & PH & 28.0047 & 0.0062 & 26.6746 & 0.0043 \\
		& PO & 1.1426 & 0.7550 & 4.3725 & 0.4536 \\
		\midrule
		GH & PH & 28.0819 & 0.0072 & 26.9723 & 0.0045 \\
		& PO & 1.7151 & 0.7208 & 4.6114 & 0.4355 \\
		\midrule
		Joe & PH & 30.1212 & 0.0061 & 28.6870 & 0.0051 \\
		& PO & 1.7149 & 0.7188 & 4.6527 & 0.4233 \\
		\bottomrule
	\end{tabular}
\end{table}

Finally, for Table \ref{tab5.8.3}, when generating from the wider YP class, all LR tests accept fitted YP models against the PH or PO regression structures, given any combination of fitted nonparametric baseline and copula function, as occurred for marginally generated data from the Weibull baseline. %This was also expected, since (nested) PH and PO models do not account for covariate short and long-term effects, but only during the whole time of follow-up.

\begin{table}[htb]
	\caption{LR tests for PH and PO classes against YP class, given fitted survival copula models over generated data from each copula model with EW YP margins ($n=500; \tau=0.25$)}
	\label{tab5.8.3}
	\centering
	%\begin{tabular}{ccrrrrrr}
	\begin{tabular}{ccrrrr}
		\toprule
		\multicolumn{1}{c}{\multirow{2}[2]{*}{Copula}} & \multicolumn{1}{c}{\multirow{2}[2]{*}{Class}} & %\multicolumn{2}{c}{Weibull Fitting} & 
		\multicolumn{2}{c}{BP Fitting} & \multicolumn{2}{c}{PE Fitting} \\
		%\cmidrule{3-8}
		\cmidrule{3-6}
		& % & LR stat. & P-value
		& LR stat. & P-value & LR stat. & P-value \\
		\midrule
		AMH & PH & 118.2559 & < 0.0001 & 111.9872 & < 0.0001 \\
		& PO & 86.1503 & < 0.0001 & 83.1002 & < 0.0001 \\
		\midrule
		Clayton & PH & 124.1241 & < 0.0001 & 117.2673 & < 0.0001 \\
		& PO & 91.4135 & < 0.0001 & 87.9554 & < 0.0001 \\
		\midrule
		Frank & PH & 118.5429 & < 0.0001 & 111.6295 & < 0.0001 \\
		& PO & 88.7532 & < 0.0001 & 84.9454 & < 0.0001 \\
		\midrule
		GH & PH & 117.8587 & < 0.0001 & 111.5238 & < 0.0001 \\
		& PO & 87.7842 & < 0.0001 & 83.9300 & < 0.0001 \\
		\midrule
		Joe & PH & 120.8868 & < 0.0001 & 114.6986 & < 0.0001 \\
		& PO & 90.1515 & < 0.0001 & 86.4063 & < 0.0001 \\
		\bottomrule
	\end{tabular}
\end{table}

Tables \ref{tab5.8.1} to \ref{tab5.8.3} showed that the analysis through the LR test for nested regression model classes, also when generating from an Archimedean survival copula with marginal EW baseline distribution (regardless of the fitted copula function or nonparametric baseline model) is a useful tool to choose the regression structure for fitting if the one that generated the data is unknown, even on a more general behavior for the true marginal baseline distribution.

\subsection{Crossing Time Estimation}

%Both PH and PO classes allows a feasible interpretation on regression parameter estimates if their corresponding assumptions are valid. However, neither of them account for the situation where, given two levels of a covariate (\emph{e.g.}, the indicator for a treatment), the associated survival functions cross each other.% at a single point of time during a follow-up.
%It might be of interest to estimate such crossing survival time $t^*$. %, given all parameter estimates for the fitted baseline distribution and regression model class (this makes sense only for Taylor Series Linearization or Parametric Bootstrap).
If it is of interest to estimate a marginal crossing survival time $t^*$, this is possible for fitted models with the YP class. However, we have seen that the standard error of $\widehat{t^*}$ has no closed form expression \citep{Dema2019} and the usual solution for it is to implement a numerical procedure to find the root that solves the nonlinear equation $S_C(t^*) - S_T(t^*) = 0$, where $S_C(\cdot)$ and $S_T(\cdot)$ are the survival functions given control and treatment values, respectively. This can be done through nonparametric bootstrap, generating a set of new samples from the original data and fitting the same model for each bootstrap sample to obtain the associated parameter estimates and the quantities $\widehat{S}_C(t^*)$ and $\widehat{S}_T(t^*)$. To search each marginal root, %for each marginal crossing survival time, %given a bootstrap sample,
the R command \texttt{uniroot} (see \citealt*{Bren1973} for more details) will be used.

% Since we did not compute the SE of the bootstrap samples for each MC replica, we will not report the ASE over the 1000 MC replicas neither the SDE of the crossing time point estimates (each one obtained by the mean over the associated bootstrap estimates).

Due to the use of a resampling method for each MC replica, the estimation of crossing survival times is far more computationally intensive. %, specially for fitted BP models. (explain and discuss the computational difficulty on BP models with high degree if required)
Thus, such evaluation %through simulation
for the survival copula models proposed here is restricted to a single correlation value, given copula generated data with Weibull YP margins and fitting only the correct copula. Consider a new MC simulation study with $M=1000$ replications of copula data sets with Weibull YP margins and $n = 500$, using the same covariates and values for baseline and regression parameters as before, but varying only the true Archimedean copula, always with a fixed $\theta$ value such that $\tau=0.25$. To estimate marginal crossing survival times associated to the treatment effect (dichotomous covariate), take two new subjects, control and treated, with covariate values $\boldsymbol{x}^*_{C;j} = (0, 0)$ and $\boldsymbol{x}^*_{T;j} = (1, 0)$, respectively, $j = 1, 2$. %Note that the continuous covariate is set constant (here, equal to a reference level) for both subjects.
%To infer on marginal crossing survival times,
Then, a nonparametric bootstrap is applied over each MC replica, using a total of $1000$ bootstrap samples to obtain their associated point and interval estimates (using the corresponding percentiles to a confidence level of 95\%).

\begin{table}[htb]
	\small
	\caption{MC statistics for marginal crossing time estimates of fitted survival copula models over generated copulas with Weibull YP margins ($n=500$; $\tau=0.25$)}
	\label{tab5.9.1}
	\centering
	\begin{tabular}{ccrrrrr}
		\toprule
		\multicolumn{1}{c}{\multirow{2}[2]{*}{Quantity}} &
		\multicolumn{1}{c}{\multirow{2}[2]{*}{Copula}} &
		\multicolumn{5}{c}{Weibull YP Fitting} \\
		\cmidrule{3-7}
		& & AE & ARB (\%) & ALB & AUB & CR (\%) \\
		\midrule
		$t_1^*$
		& AMH     & 2.2435 & 2.2842 & 1.5853 & 3.2383 & 95.4 \\ 
		& Clayton & 2.2234 & 1.3651 & 1.5865 & 3.1295 & 95.2 \\ 
		& Frank   & 2.2268 & 1.5205 & 1.5680 & 3.2149 & 95.2 \\ 
		& GH      & 2.2649 & 3.2590 & 1.5835 & 3.3060 & 94.4 \\
		& Joe     & 2.2318 & 1.7471 & 1.5932 & 3.2586 & 95.3 \\
		% \midrule
		% $t_2^*$
		% & AMH     & 1.5041 & 0.6694 & 1.2589 & 1.8630 & 95.0 \\ 
		% & Clayton & 1.4964 & 0.1520 & 1.2602 & 1.8160 & 94.7 \\ 
		% & Frank   & 1.5108 & 1.1208 & 1.2626 & 1.8814 & 95.6 \\ 
		% & GH      & 1.4963 & 0.1450 & 1.2460 & 1.8615 & 93.2 \\
		% & Joe     & 1.5034 & 0.6257 & 1.2631 & 1.8663 & 94.7 \\
		\midrule
		\multicolumn{1}{c}{\multirow{2}[2]{*}{Quantity}} &
		\multicolumn{1}{c}{\multirow{2}[2]{*}{Copula}} & \multicolumn{5}{c}{BP YP Fitting} \\
		\cmidrule{3-7}
		&& AE & ARB (\%) & ALB & AUB & CR (\%) \\
		\midrule
		$t_1^*$
		& AMH     & 2.2392 & 2.0862 & 1.5573 & 3.2609 & 96.0 \\ 
		& Clayton & 2.2237 & 1.3778 & 1.5647 & 3.1783 & 95.1 \\ 
		& Frank   & 2.2275 & 1.5521 & 1.5468 & 3.2470 & 94.2 \\ 
		& GH      & 2.2444 & 2.3242 & 1.5516 & 3.3123 & 94.7 \\
		& Joe     & 2.2212 & 1.2645 & 1.5752 & 3.2832 & 96.1 \\
		% \midrule
		% $t_2^*$
		% & AMH     & 1.5011 & 0.4668 & 1.2506 & 1.9020 & 95.1 \\ 
		% & Clayton & 1.4946 & 0.0342 & 1.2530 & 1.8560 & 94.7 \\ 
		% & Frank   & 1.5118 & 1.1868 & 1.2554 & 1.9264 & 95.1 \\ 
		% & GH      & 1.4944 & 0.0183 & 1.2391 & 1.8914 & 94.5 \\
		% & Joe     & 1.5009 & 0.4540 & 1.2560 & 1.9069 & 94.7 \\
		\midrule
		\multicolumn{1}{c}{\multirow{2}[2]{*}{Quantity}} &
		\multicolumn{1}{c}{\multirow{2}[2]{*}{Copula}} & \multicolumn{5}{c}{PE YP Fitting} \\
		\cmidrule{3-7}
		&& AE & ARB (\%) & ALB & AUB & CR (\%) \\
		\midrule
		$t_1^*$
		& AMH     & 2.2432 & 2.2674 & 1.5826 & 3.2278 & 96.3 \\ 
		& Clayton & 2.2284 & 1.5944 & 1.5880 & 3.1294 & 94.9 \\ 
		& Frank   & 2.2407 & 2.1555 & 1.5835 & 3.2242 & 94.3 \\ 
		& GH      & 2.2655 & 3.2843 & 1.5871 & 3.3091 & 94.8 \\
		& Joe     & 2.2322 & 1.7665 & 1.6068 & 3.2489 & 95.5 \\
		% \midrule
		% $t_2^*$
		% & AMH     & 1.5051 & 0.7345 & 1.2596 & 1.8560 & 94.3 \\ 
		% & Clayton & 1.4995 & 0.3626 & 1.2627 & 1.8138 & 94.3 \\ 
		% & Frank   & 1.5164 & 1.4968 & 1.2653 & 1.8810 & 95.0 \\ 
		% & GH      & 1.4955 & 0.0973 & 1.2456 & 1.8481 & 94.1 \\
		% & Joe     & 1.5006 & 0.4381 & 1.2602 & 1.8531 & 95.0 \\
		\bottomrule
	\end{tabular}
\end{table}

Table \ref{tab5.9.1} presents the MC results on the estimation of %marginal
crossing survival times on the 1st margin, divided for each copula and baseline distribution. As expected, the ARB was always lower than 4\% for the first marginal crossing survival time, %(2\% for the second one),
and the CR is at most 0.02 %(0.01)
away from the confidence level of 95\%, for %estimated crossing times over
fitted survival copula models with Weibull YP margins. %, specially for the second margin.
However, the same can be said from the estimation for fitted BP and PE models, which can still overcome the Weibull model as seen for the GH copula. Thus, such semiparametric models obtain estimates as good as those from the correctly fitted Weibull models, %even for marginal crossing survival times,
but without imposing any parametric functional form for the hazard function. Concerning the copula itself, changing only its expression (maintaining the true correlation and other unrelated parameters and quantities) has little effect over the estimated marginal crossing survival times. Also, none of them is far better or worse than another with respect to the ARB or CR values.

\section{Real Data Application}

This paper addresses the study of a manually curated data collection of patients with ovarian cancer originally described by \cite{Ganz2013}. Their resource provides data for a total of 23 distinct studies, but a single one (TCGA), with a total of $n = 508$ subjects (after removing 49 of them with missing information on the tumor stage or treatment indicators, or null values for survival times), is considered as an application to fit the proposed survival copula models. Each subject $i$, $i = 1, \ldots, n$, has 2 observed times, the first one being a nonterminal event time (in this study, the time-tumor progression) $T_i$ or a random, independent censored time $A_i$, and the second, a terminal event time (here, the overall survival of a subject) $T^*_i$ or a censored time $A_i$ (the censoring mechanism is always the same). %This is a semi-competing risks study: a subject may experience two potential events: the nonterminal event (tumor progression) or the terminal event (death). The nonterminal event may not be observable due to an earlier occurrence of the terminal event. On the other hand, an occurrence of the nonterminal event does not change the observational condition for the terminal event. Therefore, the terminal event is a competing risk for the non-terminal event, but the reciprocal is not true \citep{WuME2020}.
In order to fit a survival copula model, % it is necessary to
define %what are
the survival times and censoring statuses given $i$, at each copula margin $j$, $j = 1, 2$:
\begin{itemize}
	\item The survival time for the 1st copula margin is $Y_{i1} = \min(T_i, T^*_i, A_i)$;
	\item The censoring status for the 1st copula margin is $\delta_{i1} = \mathbb{I}(Y_{i1} = T_i)$;
	\item The survival time for the 2nd copula margin is $Y_{i2} = \min(T^*_i, A_i)$;
	\item The censoring status for the 2nd copula margin is $\delta_{i2} = \mathbb{I}(Y_{i2} = T^*_i)$.
\end{itemize}

%Defined the survival times and censoring statuses for each copula margin $j$, a set of covariates must be chosen for the regression structure specification.
From the TCGA study, two covariates are specified: \textit{CXCL12}, the concentration value of a biomarker for the gene expression of a ovarian cancer (continuous), %, with concentration values standardized to have zero mean and unit variance), 
and \textit{pltx}, an indicator for a platinum-based treatment (dichotomous, the reference level is the subject who did not receive it). Note that this %covariate 
specification is similar to the one defined for MC simulations in Chapter 5. First, all proposed survival copula models %accounting for all possible combinations of bivariate Archimedean copulas (one of AMH, Clayton, Frank, GH or Joe), baseline distributions (one among Weibull, BP or PE), and regression model classes (one of PH, PO or YP) 
are fitted in order to obtain their AIC values, which are shown in Table \ref{tab6.1}. Thus, results on the regression parameter and Kendall's $\tau$ estimates will be presented for the ``best'' combination (with respect to the AIC) of copula, baseline distribution and regression model class.

\begin{table}[htb]
	\small
	\caption{AIC values for fitted survival copula models on cancer ovarian data}
	\label{tab6.1}
	\centering	
	\begin{tabular}{crrrrrrrrr}
		\toprule
		\multicolumn{1}{c}{\multirow{2}[2]{*}{Copula}} & \multicolumn{3}{c}{PH} & \multicolumn{3}{c}{PO} & \multicolumn{3}{c}{YP} \\
		\cmidrule{2-10}
		& Weibull & BP & PE & Weibull & BP & PE & Weibull & BP & PE \\
		\midrule
		AMH     & 8621.72 & 8488.80 & 8492.15 & -- & 8485.22 & 8499.18 & -- & 8493.96 & 8482.57 \\
		Clayton & -- & 8489.49 & 8492.82 & -- & 8485.99 & 8499.25 & -- & 8494.60 & 8482.56 \\
		Frank   & -- & 8482.81 & 8485.61 & -- & 8477.73 & 8492.78 & -- & 8487.71 & 8473.49 \\
		GH      & 8602.52 & 8475.95 & 8479.39 & -- & 8471.22 & 8485.91 & -- & 8480.06 & \underline{8467.78} \\
		Joe     & -- & 8484.92 & 8487.88 & -- & 8480.05 & 8494.15 & -- & 8489.00 & 8476.54 \\
		\bottomrule
	\end{tabular}
\end{table}

For almost all combinations specifying the Weibull model for the baseline distribution, the survival copula fitting fails. Even when the log-likelihood is successfully maximized, the AIC values are much higher when compared to all fitted semiparametric models. This is an evidence of a generator process with non-monotonic hazard function for marginal survival times. All copulas are close to each other concerning the AIC criterion, but fitted models with the GH copula have the lowest AIC values, regardless of (semiparametric) baseline distribution or regression model class. Given all fitted GH models, the GH PE YP presented the lowest AIC values. %compared to all other marginal specifications for baseline and regression models.

\begin{table}[htb]
	\small
	\caption{LR tests for nested models against the GH PE YP model on cancer ovarian data}
	\label{tab6.2}
	\centering
	\begin{tabular}{crrrr}
		\toprule
		Fitted Model & Log-lik. & LR stat. & DF & P-value \\
		\midrule
		GH PE YP & -4224.89 & -- & -- & -- \\
		GH PE PH & -4234.70 & 19.62 & 4 & 5.94 $\times 10^{-4}$ \\
		GH PE PO & -4237.96 & 26.14 & 4 & 2.97 $\times 10^{-5}$ \\
		\bottomrule
	\end{tabular}
\end{table}

Now, when taking the Likelihood Ratio (LR) test statistic for the GH PE YP model against nested models %(those last ones as the null hypothesis for each test)
with the same copula and baseline specification (\emph{i.e.}, the GH PE PH and GH PE PO models) in \ref{tab6.2}, the LR statistic is significant at the level of 5\% %(1 minus the confidence level)
for both tests. Thus, there is no evidence to not reject the GH PE YP model instead of any nested model. %more parsimonious with respect to the regression model class. 
Hence, results on the regression parameter coefficients (point estimate, standard error, lower and upper limits of the 95\% confidence interval, Z-statistics and p-values) for the GH PE YP model are presented in Table \ref{tab6.3}.

\begin{table}[htb]
	\small
	\caption{Regression parameter results for the GH PE YP model on cancer ovarian data}
	\label{tab6.3}
	\centering	
	\begin{tabular}{cccrrrrrr}
		\toprule
		Margin & Covariate & Coef. & Estimate & SE & Lower & Upper & Z-stat. & P-value \\
		\midrule
		\multicolumn{1}{c}{\multirow{4}[2]{*}{1st}} & \textit{CXCL12} & $\widehat{\beta}_{11}^{(S)}$ &  0.0900 & 0.0620 & -0.0316 &  0.2116 & 1.4505 & 0.1469 \\
		& \textit{pltx} & $\widehat{\beta}_{12}^{(S)}$ & -1.7650 & 0.3898 & -2.5289 & -1.0011 & -4.5286 & 5.94 $\times 10^{-6}$ \\
		& \textit{CXCL12} & $\widehat{\beta}_{11}^{(L)}$ &  1.8903 & 0.7537 & 0.4131 & 3.3675 & 2.5081 & 0.0121 \\
		& \textit{pltx} & $\widehat{\beta}_{12}^{(L)}$ & 8.1402 & 14.1470 & -19.5874 & 35.8678 & 0.5754 & 0.5650 \\
		\midrule
		\multicolumn{1}{c}{\multirow{4}[2]{*}{2nd}} & \textit{CXCL12} & $\widehat{\beta}_{21}^{(S)}$ & 0.2372 & 0.1087 &  0.0241 & 0.4503 & 2.1817 & 0.0291 \\
		& \textit{pltx} & $\widehat{\beta}_{22}^{(S)}$ & -1.2938 & 0.4460 & -2.1680 & -0.4196 & -2.9007 & 0.0037 \\
		& \textit{CXCL12} & $\widehat{\beta}_{21}^{(L)}$ & -0.1160 &  0.1414 & -0.3932 & 0.1612 & -0.8205 & 0.4119 \\
		& \textit{pltx} & $\widehat{\beta}_{22}^{(L)}$ & -0.4670 &  0.6098 & -1.6621 & 0.7281 & -0.7659 & 0.4437 \\
		\bottomrule
	\end{tabular}
\end{table}

Given the GH PE YP model in Table \ref{tab6.3}, except for $\widehat{\beta}_{11}^{(S)}$, all other short-term regression coefficients were significant at the level of 5\%. However, the long-term counterpart of $\widehat{\beta}_{11}^{(S)}$, $\widehat{\beta}_{11}^{(L)}$, has significance. %Since the PO model arises as a particular case from the YP model if all long-term parameters are supposed to be null, ignoring the potential significance of any long-term parameter can affect conclusions for the erstwhile short-term parameters, not only on their significance, but also on their magnitude.
When interpreting the significant regression coefficients for the GH PE YP model, it can be said that the ratio of hazard rates between a treated subject and a control for tumor progression (nonterminal event) is $\exp(-1.7650) \approx 0.1712$ (or $17.12\%$), \emph{i.e.}, the treatment (\textit{pltx}) reduces the hazard rate in $82.88\%$ for a tumor progression. Also, the ratio of hazard rates between a treated subject and a control for death (terminal event) is $\exp(-1.2938) \approx 0.2742$ (or $27.42\%$). In other words, the treatment reduces the hazard rate in $72.58\%$ for a death. On the other hand, each level gained for the \textit{CXCL12} biomarker increases the hazard rate in $\exp(1.8903) \approx 6.6214$ times in the long-term for a tumor progression, and in $\exp(0.2372) \approx 1.2677$ times in the short-term for a death. Therefore, the platinum treatment reduces the hazard for both events in the short-term, but does not have significant influence in the long-term. However, greater levels of \textit{CXCL12} increases the hazard for both events, but in distinct moments of the follow-up. Concerning the dependence estimation, for the fitted GH PE YP survival copula model, the estimated Kendall's correlation is $\widehat{\tau} = 0.3332$, with confidence interval $I_{\widehat{\tau}} = [0.1576, 0.4483]$. %This significant correlation gives an estimated upper tail dependence $\widehat{\chi}_U = 2 - 2^{1/0.3332} \approx -6.0067$. In other words, marginal survival times have moderate overall dependence given a same subject, but they also have a moderate upper tail dependence, thus implying on a mild correlation across smaller survival times. For greater marginal survival times, there is no lower tail dependence, since it is always null for GH copulas.

\section{Conclusions}

The present paper proposes a fully likelihood-based approach for modeling on multivariate clustered survival data, under independent random censoring, by introducing one of five Archimedean copulas (AMH, Clayton, Frank, GH or Joe), and marginally specifying one of three baseline distributions (Weibull, BP or PE) and the YP regression model class, or one of its particular cases (PH or PO). The main differences with respect to other works on survival copula modeling are: the marginal semiparametric model fitting (BP and PE), which allows a much more flexible representation of the baseline hazard or odds function, and the YP class to account for potentially distinct covariate effects on short and long terms, along with marginal crossing survival times. These compound the main contribution of the paper, since other studies %so far
on marginal semiparametric YP modeling for survival Archimedean copulas cannot be found in the literature. %Besides, it is relatively easy to maximize the log-likelihood through numerical routines in order to find marginal parameter and crossing survival time estimates.

To evaluate the proposed modeling, an extensive study on simulated data was realized. For that study, all data was generated from a bivariate survival copula model with two covariates for each margin. %The generation could vary on the copula used (one of the five Archimedean copulas aforementioned); on the survival model for the baseline distribution on both margins (from Weibull or EW models), or on the regression model class (PH, PO or YP). For a fixed sample size $n=500$, simulation scenarios also were divided by three chosen values for the true Kendall's $\tau$ correlation ($\tau \in \{0.25, 0.5, 0.75\}$). In its turn, the proposed survival copula models for fitting over simulated data were always a combination of a copula function (again, one of the five Archimedean copulas), a baseline distribution (one of Weibull, BP or PE models), and a regression model class (one of PH, PO or YP classes). 
Fitted models were compared by exchanging the copula function or baseline distribution for parameter estimation and choice through the mean AIC,while for an analysis through the LR test they were compared by swapping the regression model class.
%Finally, all proposed survival copula models for fitting were applied to a set ...
A set of bivariate real data was also used, in order to choose the best combination of fitted copula, baseline distribution and regression model class by the AIC value combined to a LR test analysis, and also present its results on regression parameter and Kendall's $\tau$ correlation estimates.

One of the main goals of this work was to compare results on regression parameter estimates %, specially in terms of the Average Relative Bias (ARB) and Coverage Rate (CR),
and mean AIC among each Archimedean copula used for fitting, given a fixed copula for generation, %of the simulated data,
and verify if a correct copula fitting has suitable results under a high sample size for three distinct levels of dependence. Illustrating for the AMH copula, %in the main text and presenting results when generating from other copulas in the Appendix,
fitting the correct copula has generally produced good estimates (when looking for the ARB and CR), although for many cases a wrong fitting of the copula model can still have similar performance, even for a scenario with higher correlation. In general, fitting a wrong copula does not lead to a severe loss of performance for the regression parameter estimation, but this in fact occurs for the computed AIC and $\tau$ estimates.

Given only fitted models with the correct copula and regression structures, another main goal of this work was to compare results for three baseline distributions, two of them offering a nonparametric appeal. As expected, when generating copula data with Weibull margins, fitting the same baseline yielded the best results, although fitted models with a nonparametric baseline (BP or PE) follow closely in terms of ARB. On the other hand, when generating copula data with EW margins, PE and BP models perform better than Weibull ones for the majority of copulas. This was expected due to their nonparametric nature: their number of baseline parameters depends on the sample size, thus fitting well when that size is high, including when the true marginal hazard functions have a non-monotonic behavior.

Finally, given the correct copula fitting and now exchanging the regression model class, the last goal involved an analysis through the LR test to check if the YP class is preferable or no over nested PH and PO classes. Given one of the smaller regression structures for marginal survival data generation, the YP class fits significantly better if tested against the other not used for generation, but it does not if tested against the true one, regardless of the true or fitted baseline distribution and copula function. On the other hand, if the YP class is also part of the generator process for marginal survival data, LR tests always accept it against both PH and PO classes.

As supplement to this paper, we provide an Shiny application to show results for any scenario not presented in the main text, including boxplots for relative bias on parameter estimates, AIC values and marginal crossing survival times. The application is available at \url{https://wrmfstat.shinyapps.io/CopRegEst/}. For future research, we intend to develop a Bayesian approach for the proposed modeling presented here. We also suggest that it can be extended to incorporate cure fraction estimation when the marginal survival function seems to stop at a positive lower limit. Another possibility of extension is to extend for a joint frailty-copula modeling, when the heterogeneity can come from a known source other than the clustering among subjects themselves, such as interviews from a small number of distinct studies (see \citealt*{WuME2020} for an example).

\bibliographystyle{unsrtnat}
\bibliography{references}  %%% Uncomment this line and comment out the ``thebibliography'' section below to use the external .bib file (using bibtex) .



% \end{thebibliography}

\end{document}